\documentclass[aps,prl,twocolumn,superscriptaddress, longbibliography, 10pt]{revtex4-2}
\usepackage{graphicx} 
\usepackage[caption=false]{subfig}

\usepackage{amsmath,amssymb}
\usepackage{ulem}
\usepackage{bm,hyperref}
\usepackage[dvipsnames]{xcolor}

\newcommand{\pheliqs}{Univ. Grenoble Alpes, CEA, Grenoble INP, IRIG, PHELIQS, F-38000 Grenoble, France}
\newcommand{\neel}{Univ. Grenoble Alpes, CNRS, Institut Néel, F-38000 Grenoble, France}
\newcommand{\lncmiG}{Univ. Grenoble Alpes, INSA Toulouse, Univ. Toulouse Paul Sabatier, EMFL, CNRS, LNCMI, F-38000 Grenoble, France}
\newcommand{\imr}{Institute for Materials Research, Tohoku University, Oarai, Ibaraki, 311-1313, Japan}
\usepackage[final]{changes}
\definechangesauthor[name = Georg, color=blue]{GK}
\definechangesauthor[name = Daniel, color=green]{D}
\definechangesauthor[name = Jean-Pascal, color=magenta]{JP}
\definechangesauthor[name = Timothee, color=cyan]{TV}

\begin{document}

\title{Connecting High-Field and High-Pressure Superconductivity in \texorpdfstring{UTe$_2$}{UTe2}}

\author{T.~Vasina}
\affiliation{\pheliqs}
\author{D.~Aoki}
\affiliation{\imr}
\author{A.~Miyake}
\affiliation{\imr}
\author{G.~Seyfarth}
\affiliation{\lncmiG}
\author{A.~Pourret}
\affiliation{\pheliqs}
\author{C.~Marcenat}
\affiliation{\pheliqs}
\author{M.~Amano Patino}
\affiliation{\pheliqs}
\affiliation{\neel}
\author{G.~Lapertot}
\affiliation{\pheliqs}
\author{J.~Flouquet}
\affiliation{\pheliqs}
\author{J.-P.~Brison}
\affiliation{\pheliqs}
\author{D.~Braithwaite}
\email[E-mail me at: ]{daniel.braithwaite@cea.fr}
\affiliation{\pheliqs}
\author{G.~Knebel}
\email[E-mail me at: ]{georg.knebel@cea.fr}
\affiliation{\pheliqs}

\date{\today}

\begin{abstract}
   The existence of multiple superconducting phases induced by either pressure or magnetic field is one of the most striking features of superconductivity of UTe$_2$, among the many unusual superconducting properties of this system. Here we report thermodynamic measurements of the superconducting phase diagram combining pressure and magnetic fields up to $30$~T. We show that the ambient pressure, high-field, superconducting phase evolves continuously with pressure to join the high-pressure, zero-field superconducting phase. This proves that these two phases are one and the same, and must have the same order parameter.   
\end{abstract}

\maketitle
One of the most singular aspects  of unconventional superconductivity (SC) is that a purely electronic pairing mechanism allows a multitude of possible superconducting pairing states. This opens the possibility of stabilizing several different superconducting states in the same material by varying external parameters like temperature, magnetic field and pressure. 
There are now hundreds of known unconventional superconductors in the different families of heavy fermions, organics, high $T_c$ cuprates, iron-based pnictides, or Kagome superconductors \cite{Stewart2017, Jiang2022}. 
However, only very few are known to display multiple superconducting states \cite{Leggett1975, Fisher1989, Hasselbach1989, Adenwalla1990, Ott1985, Khim2021}. 
The recently discovered spin-triplet candidate superconductor UTe$_2$ \cite{Ran2019, Aoki2019} is most intriguing as the field- or pressure-induced phases have a higher critical temperature than the superconducting ground state \cite{Aoki2022review, Lewin2023}. 
It was first shown that when a small hydrostatic pressure is applied, two distinct superconducting transitions occur versus temperature \cite{Braithwaite2019, Thomas2020}.  
But the most remarkable phenomenon in UTe$_2$ with orthorhonbic crystal structure is undoubtedly the effect of an applied magnetic field on its superconducting properties \cite{Ran2019a, Knebel2019, Aoki2022review, Lewin2023}. 
In particular, when the field is applied along the magnetic hard $b$-axis, a spectacular reinforcement of SC is found above $15$~T up to a field of about $35$~T, where a first-order metamagnetic transition to a strongly polarized state occurs \cite{Knafo2019, Knebel2019, Ran2019a}. 

Precise calorimetry measurements in high magnetic field recently showed that there is a clear phase transition between a low-field (LF) and a high-field (HF) superconducting state \cite{Rosuel2023}. 
The HF phase cannot be 
soly explained by a symmetry change as in CeRh$_2$As$_2$ \cite{Khim2021} or a rotation of the $\boldmath{d}$-vector  
of the spin-triplet superconducting order parameter as in UPt$_3$ \cite{Sauls1994}. It also requires an enhancement of the pairing strength, associated with the emergence of a new pairing mechanism, apparently  driven by the fluctuations associated to the metamagnetic transition \cite{Rosuel2023, Tokunaga2023}. 

\begin{figure}[bt]
	\includegraphics[width=\linewidth]{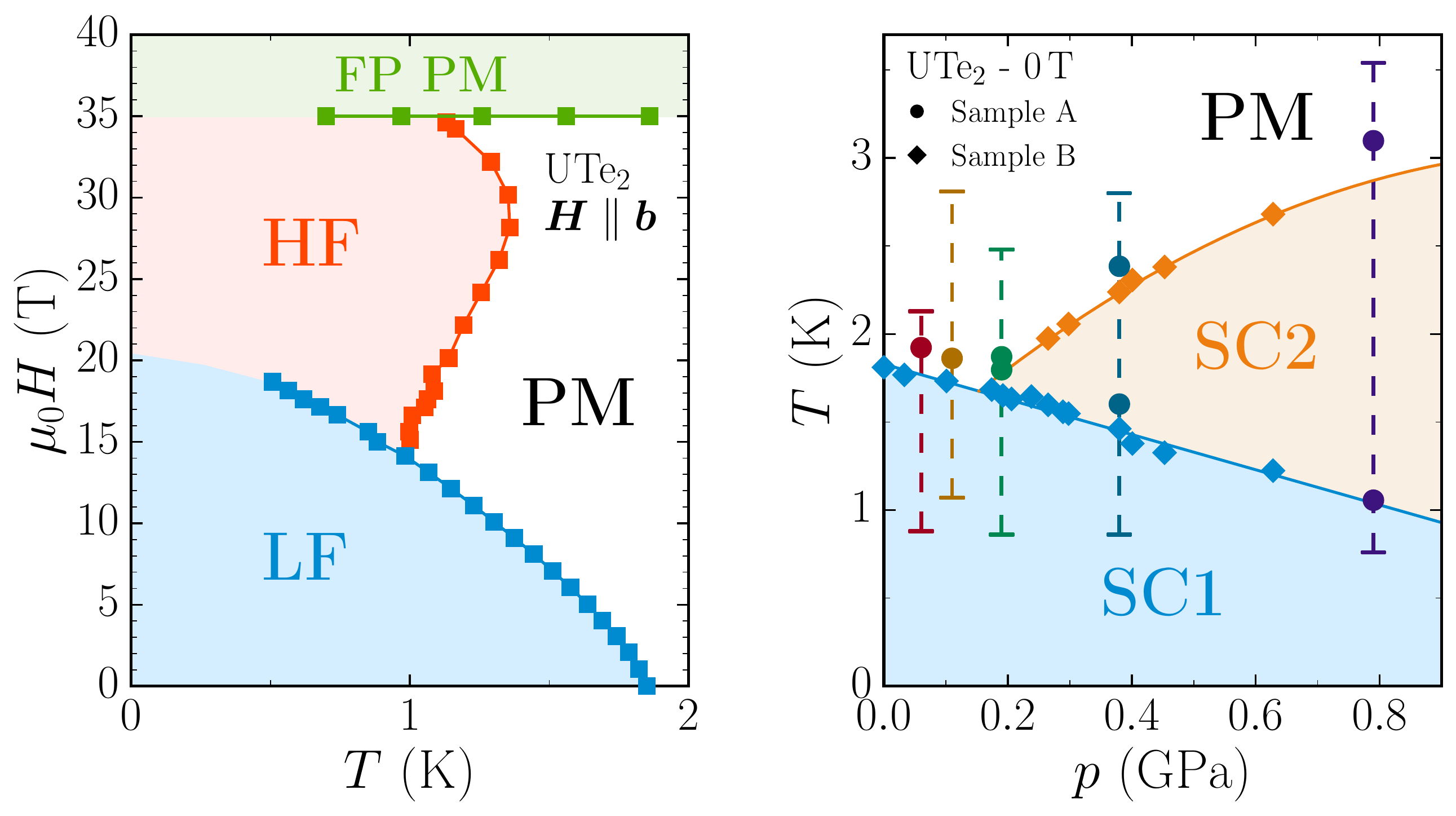}
	\caption{The phase diagrams of multiple superconducting phases in UTe$_2$ at ambient pressure under magnetic field $H \parallel b$ (left) adapted from \cite{Rosuel2023}, confirmed in \cite{Kinjo2023PRB, Sakai2023} and under pressure in zero field (right). The right-hand figure shows points from two of the three studies reported here (Samples A and B), and is similar to previous reports \cite{Braithwaite2019, Aoki2020, Lin2020, Thomas2020}. The dashed vertical lines show the pressures and temperature ranges for the study on sample A measured up to $30$~T presented here.}
	\label{Fig01}
\end{figure}

\begin{figure*}[ht]
	\includegraphics[width=0.9\linewidth]{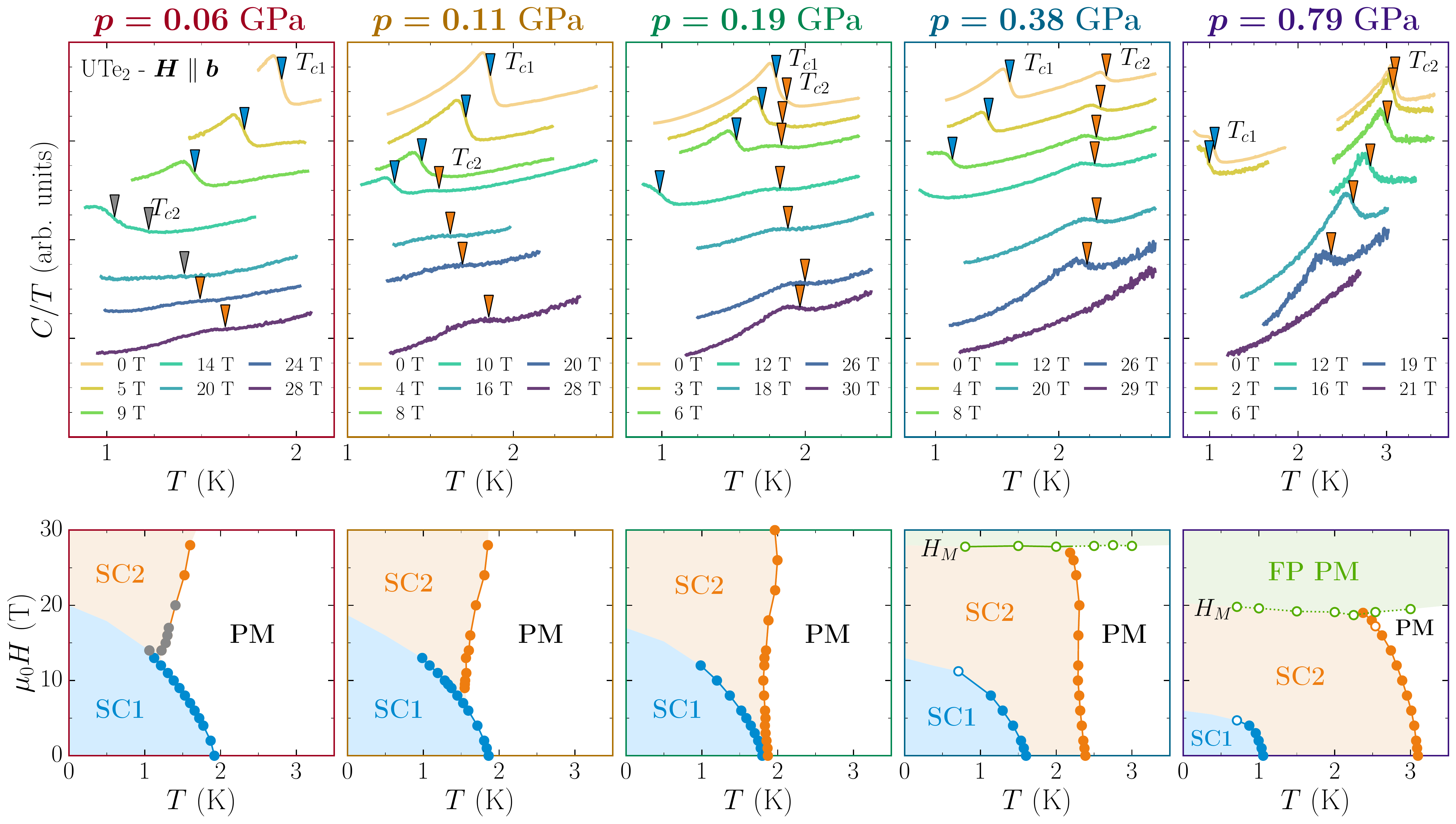}
	\caption{Top: $C/T$ curves at selected fields for sample A at the five measured pressures. Blue and orange arrows show the position of $T_{c1}$ and $T_{c2}$ respectively. The data are shifted vertically for clarity. The positions of $T_{c1}$ and $T_{c2}$ are extracted from the fit using a Gaussian distribution of critical temperatures, except for some points where the overlap of the transitions was too large to apply this procedure (points in grey). In that case, $T_{c1}$ and $T_{c2}$ are placed by eye taking the midpoint of the transition in the signal phase. Bottom: Evolution of the phase diagram with pressure showing that the high-field phase SC2 evolves continuously into the high-pressure phase. Closed (open) circles for the superconducting phase boundaries are extracted from temperature (field) sweeps. Values of the metamagnetic field $H_M$ are extracted from field sweeps.}
	\label{Fig02}
\end{figure*}

Figure \ref{Fig01} shows the phase diagrams of these superconducting phases at ambient pressure for a field $H \parallel b$-axis, and under hydrostatic pressure $p$ at zero field. In both cases only two superconducting phases are shown, although
general thermodynamic considerations \cite{LandauLifshitzVol5} suggest that a third phase should be present in both the high-field and the high-pressure phase diagram.  
To date, it has not been unambiguously detected experimentally.
Only in the superconducting phase at ambient pressure transport measurements suggest such an additional line \cite{Sakai2023, Tokiwa2023}, which remains elusive for thermodynamic probes.
Characterizing and identifying these different superconducting phases is now a priority for UTe$_2$ as it is expected that some or all of them have a spin-triplet order parameter, highly sought after for its original fundamental properties as well as its interest in quantum engineering \cite{Sato2017}. 
Even the superconducting order parameter symmetry of the LF phase at ambient pressure has not been identified unambiguously \cite{Bae2021, Matsumura2023, Ishihara2023, Iguchi2023, Theuss2024, Hayes2024}.
Obviously in Fig.~\ref{Fig01} the LF phase is the same as the SC1 phase under pressure. Now that there is proof of the existence of two distinct superconducting phases with applied field, the link between the HF and SC2 phases must be clarified.

To resolve this question, we have performed calorimetry measurements under pressure in high magnetic fields $H \parallel b$-axis on single crystals of UTe$_2$, in order to get a thermodynamic determination of the different phases and their precise evolution with combined pressure and field.
Three different samples have been investigated, details on the samples and experiments are given in the Supplementary Material \cite{SuppMat}.

The main result of this work is shown in Fig.~\ref{Fig02}. Sample A was measured at five pressures, whose positions in the $(p,T)$ phase diagram are indicated by vertical dashed lines in Fig.~\ref{Fig01}. 
The top graphs in Fig.~\ref{Fig02} show $C/T$ versus $T$ for some selected magnetic fields (additional data are shown in the Supplemental Material \cite{SuppMat}). 
The lower graphs indicate the corresponding phase diagrams. 
For $p=0.06$~GPa, the overall aspect of the curves and the obtained phase diagram are quite similar to those obtained at $p=0$ in the previous study \cite{Rosuel2023}. 
At zero field only one transition at $T_{c1}$ is present, but for $H \gtrsim 14$~T, the second transition at $T_{c2}$ is observed as a clear but very broad anomaly in $C/T$ vs $T$. 
Because of its broadness, we analyzed the data  
assuming a Gaussian distribution of critical temperatures  
similar to Rosuel et al. in the ambient pressure study \cite{Rosuel2023}, in order to extract the critical temperature and characterize the width and amplitude of the specific heat jump. 
Details on the analysis are given in the Supplemental Material \cite{SuppMat}.  
At $p=0.11$~GPa, it is apparent that the low-field phase has shrunk slightly, and the high-field phase has expanded to higher temperatures and lower fields. 
Indeed, the high-field transition now appears for $H \approx 10$~T. 
At $0.19$~GPa both transitions start to be visible at zero field. 
The high-field phase has further expanded and now encompasses completely the low-field phase. 
At $p= 0.38$ and $0.79$~GPa the low-field phase continues to shrink to lower temperatures and fields.  
While for $0.38$~GPa the upper critical field $H_{c2}$ is almost vertical with a shallow  "S"-shape, for $0.79$~GPa it has the usual curvature up to $H_M$.

The continuous evolution of the superconducting phase diagrams at low pressures rigorously
demonstrates that the ambient-pressure/high-field (HF in Fig.~\ref{Fig01}) and high-pressure/zero-field (SC2 in Fig.~\ref{Fig01}) superconducting phases are identical
and should be characterized by a superconducting order parameter having the same symmetries. 
Hence, we have labelled this phase SC2 in both cases.

\begin{figure}[bt]
	\includegraphics[width=0.9\linewidth]{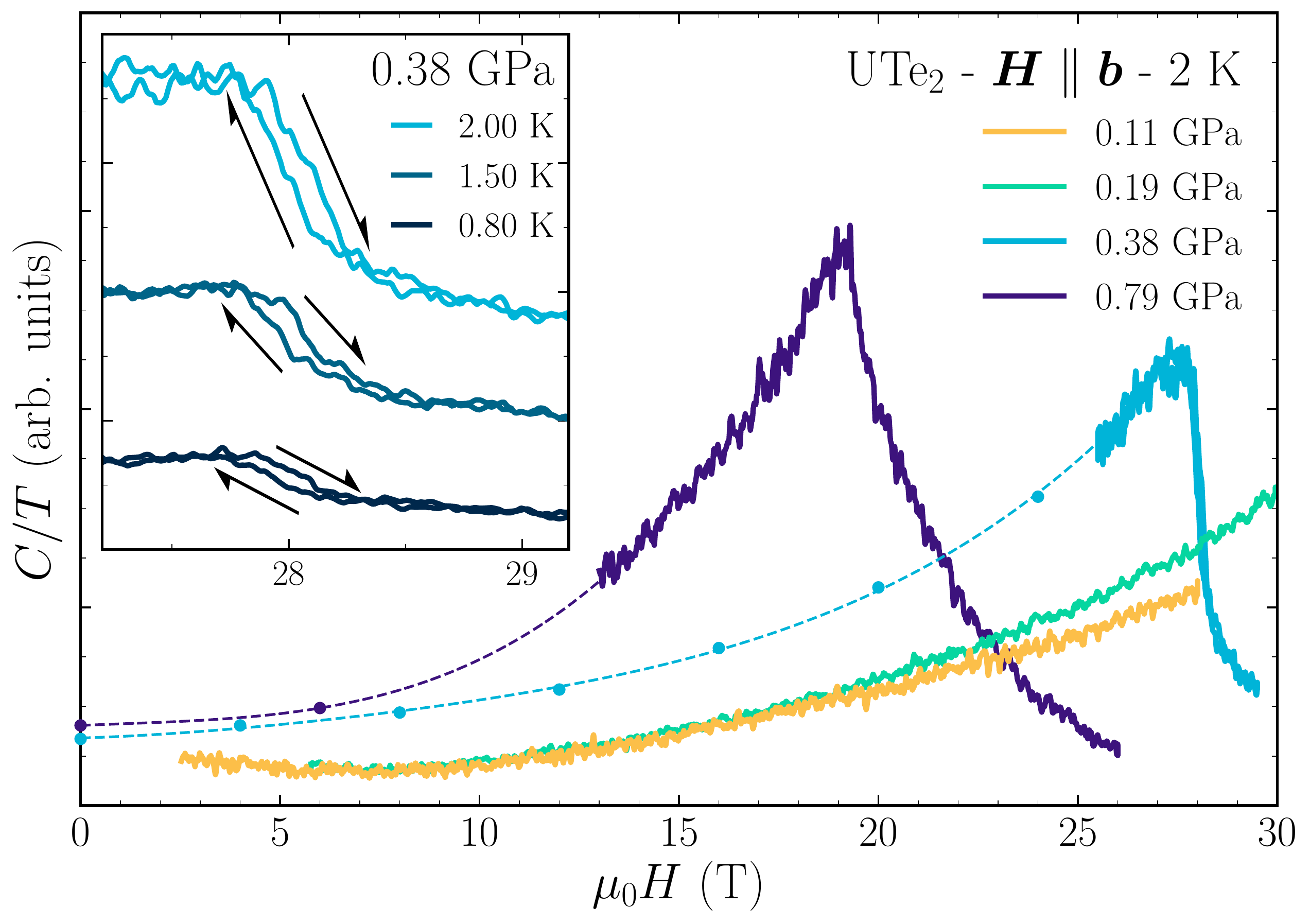}
	\caption{Field sweeps for different pressure at $2$~K on sample A. Continuous sweeps were performed for a partial field range (continuous line). The discontinuous points (circles and dashed lines) were extracted from temperature sweeps. The inset shows the temperature dependence of the hysteresis observed at $0.38$~GPa.}
	\label{Fig03}
\end{figure}

For the first three pressures, the metamagnetic transition field, $H_M$, reaching $35$~T at ambient pressure, is at values above the highest field ($30$~T) available for these pressure studies. As expected \cite{Lin2020, Knebel2020}, $H_M$ decreases with pressure and at $0.38$~GPa and above we can observe this first-order transition that continues to act as a cut-off for SC. 
Figure \ref{Fig03} displays field sweeps of $C/T$ at constant temperature. 
This allows us to determine $H_M$, once its value decreases below $30$~T under the effect of pressure. 
It also allows us to characterize the variation of $C/T$ as we cross $H_M$. 
At $0.38$~GPa, the lowest pressure where we could reach $H_M$, the shape of the transition is quite similar to that observed at ambient pressure \cite{Rosuel2023}. 
At $2$~K (in the normal state), $C/T$ continuously increases with field on approaching $H_M$, followed by a sharp, almost step-like, decrease of $C/T$ at $H_M$, on the high-field side of the transition. 
We also observed a small hysteresis at $H_M$
, indicating the first-order nature of the transition. 
This hysteresis disappears for temperatures above $2.5$~K implying that the temperature of the critical point of the first-order transition decreases with pressure, as observed in a previous study of resistivity under pressure \cite{Knebel2024}. At $0.79$~GPa the transition at $H_M$ evolves into an almost symmetrical peak, and no hysteresis is observed. 
This suggests that at this pressure, the transition into the polarized state becomes second-order or a sharp cross-over rather than a first-order transition, even at the lowest temperatures. 
Nevertheless, it continues to act as a cut-off for the superconducting state (see Fig.~\ref{Fig02}).

Although our measurement is not quantitative, we can also get some information on the evolution of $(C/T)_n$, the specific heat coefficient in the normal phase (above the max of $T_{c1}$ and $T_{c2}$), with pressure and field. 
It is known that at ambient pressure and above $\approx$ $15$~T, $(C/T)_n$ increases strongly with field up to $H_M$ \cite{Rosuel2023,Imajo2019, Miyake2021b}. This increase most likely corresponds to an increase of the effective electronic mass $m^*$, connected to the enhancement of the superconducting pairing strength as field is increased towards $H_M$. A similar increase of $m^*$ is also observed in electrical transport \cite{Knebel2019, Knafo2019, Knafo2020} and in the magnetic fluctuations detected by NMR \cite{Tokunaga2023}, 
As pressure is initially increased, $(C/T)_n$ logically rises at a faster rate, as $H_M$ is shifted to lower fields. 
However, the curve at $0.79$~GPa on Fig.~\ref{Fig03} shows that $(C/T)_n$ seems to reach higher absolute values at $H_M$ as pressure increases. The ac calorimetry technique under pressure is not sufficiently precise to distinguish whether this is merely due to the fact that increasing pressure already enhances $m^*$ at zero field, or whether the field-induced enhancement is actually stronger under pressure. Nevertheless, it confirms that the absolute values of $m^*$ on approaching $H_M$ do become larger with pressure. 

\begin{figure}[t]
\includegraphics[width=0.9\linewidth]{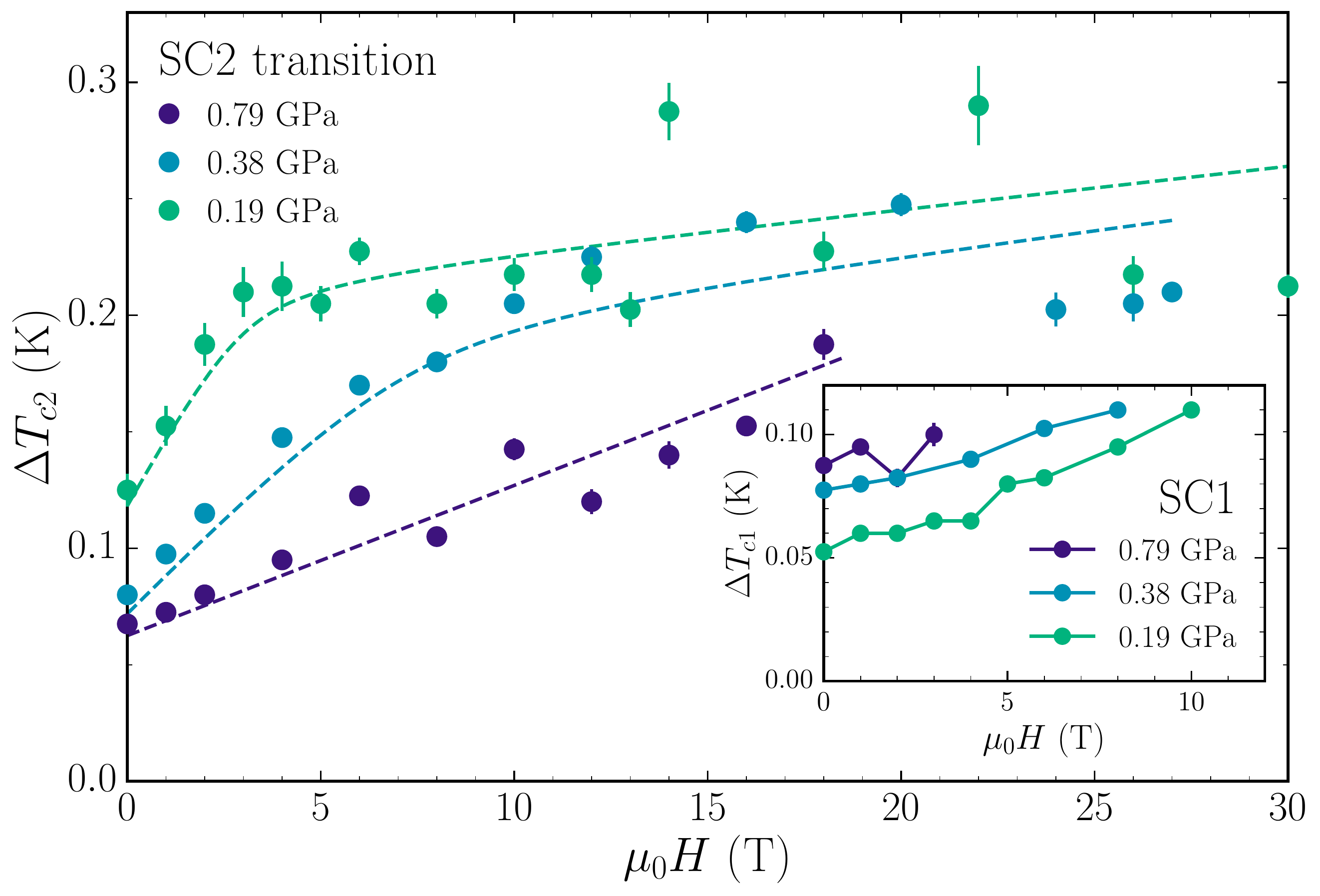}
\caption{Widths $\Delta T_{c2}$ of the transition at SC2 as a function of magnetic field, extracted by fitting a Gaussian distribution of critical temperatures \cite{SuppMat}. The inset shows the width of the SC1 transition. (Dashed lines are guides for the eye.)}
\label{Fig04}
\end{figure}

Another characteristic feature of the SC2 phase in high field is the extremely broad transition compared to the sharp transition at low field, as previously observed in the ambient pressure study \cite{Rosuel2023}. The transition widths $\Delta T_{c2}$ versus field for different pressures as deduced from the Gaussian analysis of the critical temperature distribution are shown in Fig.~\ref{Fig04}. There is some scatter due to the noise in the high-field data, but the general trend is quite clear. 
 Under pressure the transition for SC2 becomes sharp at low field, with a width similar to that of the SC1 transition. However the width increases strongly with field, and seems to tend towards a value quite similar to the width in the SC2 phase at ambient pressure. For comparison we also show the width of the SC1 transition, which is at least twice as sharp at the highest fields where it can be determined than that of the SC2 transition at high fields. So it appears that the broad transition at SC2 in high field is a robust feature of UTe$_2$, persisting under pressure even when the transition is quite sharp at low field. 
 This broadening is most likely deeply connected to the mechanism driving this superconducting phase. 
 
 \begin{figure}[tb]
\includegraphics[width=0.9\linewidth]{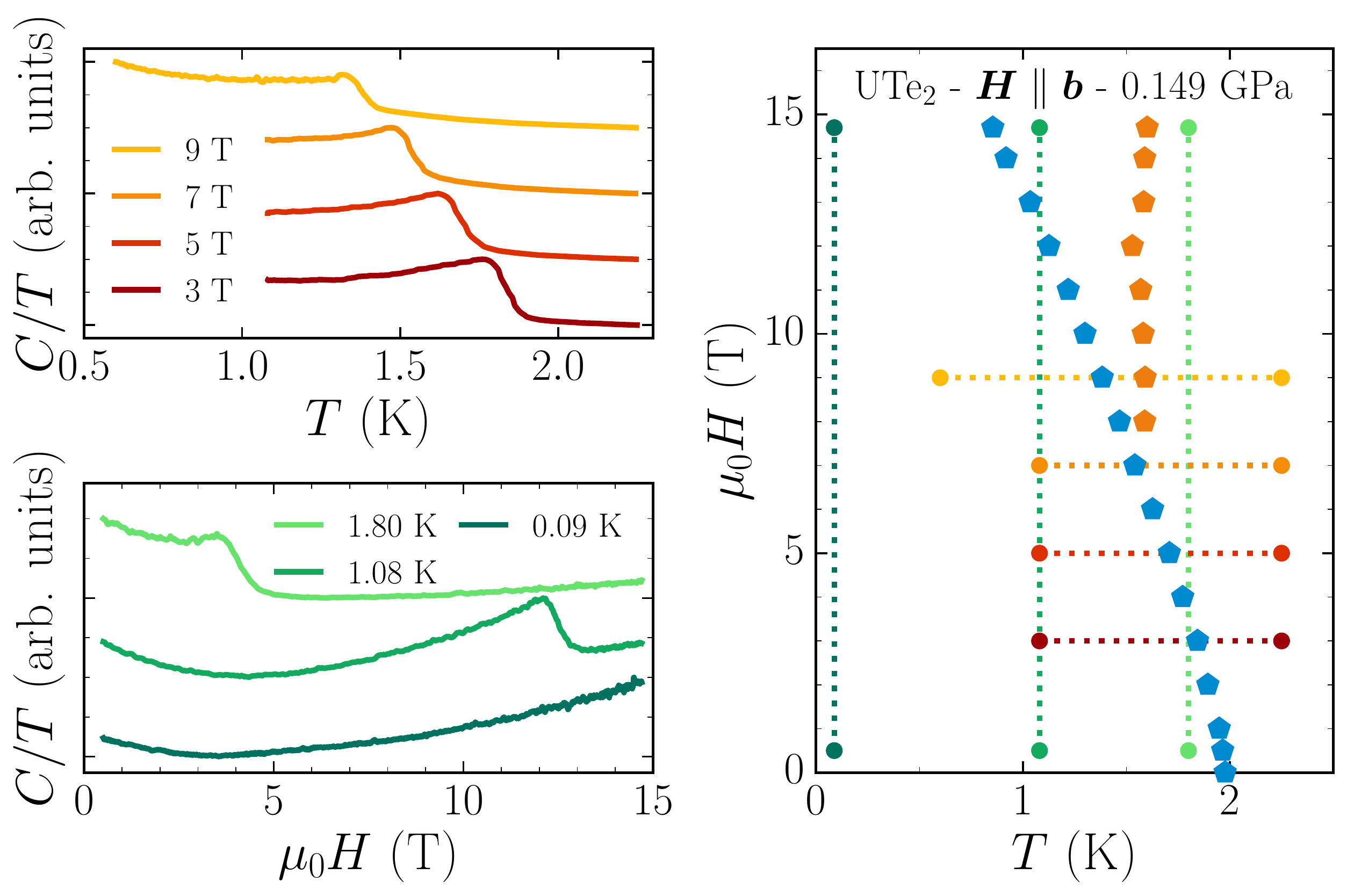}
\caption{(top left) Temperature sweeps at different field, measured on Sample C at $0.149$~GPa. (bottom left) Field sweeps measured in the same conditions, at different temperatures. (right) $(T,H)$ phase diagram at $0.149$~GPa. Blue (orange) pentagons represent SC1 (SC2) transitions. The colored dotted lines summarize the sweeps shown on the left.} 
\label{fig:fourth}
\end{figure}

As already evoked in the introduction, an 
important question concerning the  phase diagram is how the phase boundary lines of SC1 and SC2 meet, as three second-order transition lines meeting in one point is generally not thermodynamically allowed \cite{LandauLifshitzVol5, Yip1991}. \added[id=GK]{In order to allow the crossing of three lines, special restrictions are needed: two lines should have the same slope while the third line exhibits a specific heat jump that vanishes towards the critical point (see Suppl. Material of Ref.~\cite{Braithwaite2019}). }
An experimental difficulty is that the broad transition of the SC2 phase does not allow to follow accurately $T_{c1}$ and $T_{c2}$ \added[id = GK]{and their jumps $\Delta C$} when both transitions overlap. Figure~\ref{fig:fourth} illustrates our strategy to find a transition line inside SC1 in the $(T,H)$ phase diagram, at an intermediate pressure of $0.149$~GPa for sample C. Both temperature and field scans were performed at this pressure, as depicted by colored dotted lines in the $(T,H)$ phase diagram on the right. The specific heat data extracted from these measurements is shown on the left. 
No sign of any fourth transition line could be detected inside the SC1 phase, neither by temperature or field sweeps. This question is also open for the $(p,T)$ phase diagram [see Fig.~1 (right)], where it is even harder to resolve, as measurements cannot be performed while continuously sweeping pressure. Only a precise quantitative measurement of $C/T$ vs $T$ for different pressures might reveal a change of the superconducting state and suggest a vertical line in the phase diagram, but this is beyond the precision of our study.

The present work is the first to demonstrate the continuous evolution of the SC2 phase in the $(T,H,p)$ phase diagram. 
However, other experiments had already probed the different phases across this phase diagram. 
Notably tunnel diode oscillator measurements \cite{Lin2020} had shown that in the high-pressure $(T,H)$ phase diagrams, the low-temperature superconducting phase is completely encompassed inside the high-temperature superconducting phase. More recently,
NMR spectroscopy performed both at ambient pressure and high field, or at low field under pressure, shows no drop of the Knight shift when going from the normal state to SC2 \cite{Kinjo2023PRB, Kinjo2023}, leading to the conclusion that the SC2 phase would be spin-triplet. 
 This is indeed a very strong argument for a spin-triplet state, notably because by contrast at lower field and zero pressure, a decrease of the Knight shift is detected along the same field direction \cite{Kinjo2023PRB}.
A slight uneasiness with this conclusion is that at ambient pressure, the SC2 phase is probed in high fields, where the decrease of the Knight shift should be anyway strongly suppressed, whereas under pressure, the temperature-dependent background is more difficult to evaluate with certainty than at ambient pressure. 
Compared with the calorimetry results, 
the strong broadening of the specific heat anomaly along the SC2 boundary at ambient pressure (including its angular dependence) could be
quantitatively explained from a spread of the values of $H_M$ if SC2 is spin-singlet \cite{Rosuel2023}. 
The persistence of this anomalous high-field broadening under pressure revealed in this work, shows that this is clearly a characteristic feature of the SC2 phase, which cannot be easily put aside.

From a theoretical point of view, it has also be shown that in UTe$_2$, many different superconducting states seems to be competing, as well as different pairing mechanisms connected to antiferromagnetic or ferromagnetic fluctuations \cite{Xu2019, Shishidou2021, Ishizuka2021, KreiselPRB2022, TeiPRB2024}. 
No model can presently account for the complex experimental situation under pressure where valence \cite{Wilhelm2023} is slowly changing as well as magnetocrystalline anisotropy \cite{Li2021} and most likely also the Fermi surface. 
However, recent work \cite{HakunoPRB2024} showed that antiferromagnetic fluctuations (with wave vector along the $b$-axis) and ferromagnetic correlations along the uranium chains ($a$-axis) could cooperate to promote a spin-triplet state. 
By contrast, the development of a component of the antiferromagnetic wave vector along the $a$-axis would suppress the ferromagnetic fluctuations and promote either a spin-singlet or a different (B$_{2u}$) spin-triplet superconducting state.
As the antiferromagnetic long-range order above $1.5$~GPa is precisely characterized by a wave vector \cite{Knafo2023} differing from that of the ambient pressure fluctuations \cite{Duan2020, Knafo2021} by a small component along the $a$-axis, this is indeed an appealing scenario, independently of the precise parameters controlling this evolution under pressure.

Our rigorous demonstration by a bulk thermodynamic probe that the high-field and high-pressure superconducting phases are one and the same, clarifies the complex superconducting phase diagram of UTe$_2$. 
Presently, they do not allow for a more precise determination of the symmetry of the order parameter in the SC2 phase. However, these results open a new theoretical approach to the question of the mechanisms and symmetries characterizing the SC2 phase: 
the proposed scenarios should apply as well for the HF phase seemingly triggered by the approach to $H_M$, 
and for the “high-pressure” phase most likely emerging together with stronger antiferromagnetic fluctuations. 
This should both add guides and constraints to the models, and also enhances the possibilities to experimentally probe these scenarios.

\begin{acknowledgments}
We are particularly grateful for many stimulating discussions with Y. Yanase, M. Zhitomirsky, S. Fujimoto, K. Ishida. 
We acknowledge financial support from the French National Agency for Research ANR within the project FRESCO No. ANR-20-CE30-0020, SCATE No. ANR-22-CE30-0040 and from the JSPS programs KAKENHI (JP19H00646, JP20K20889, JP20H00130, JP20KK0061, JP20K03854, JP22H04933). We acknowledge support of the LNCMI-CNRS, member of the European Magnetic Field Laboratory (EMFL), and from the Laboratoire d’excellence LANEF (ANR-10-LABX-0051). 
\end{acknowledgments}


\begin{thebibliography}{57}%
	\makeatletter
	\providecommand \@ifxundefined [1]{%
		\@ifx{#1\undefined}
	}%
	\providecommand \@ifnum [1]{%
		\ifnum #1\expandafter \@firstoftwo
		\else \expandafter \@secondoftwo
		\fi
	}%
	\providecommand \@ifx [1]{%
		\ifx #1\expandafter \@firstoftwo
		\else \expandafter \@secondoftwo
		\fi
	}%
	\providecommand \natexlab [1]{#1}%
	\providecommand \enquote  [1]{``#1''}%
	\providecommand \bibnamefont  [1]{#1}%
	\providecommand \bibfnamefont [1]{#1}%
	\providecommand \citenamefont [1]{#1}%
	\providecommand \href@noop [0]{\@secondoftwo}%
	\providecommand \href [0]{\begingroup \@sanitize@url \@href}%
	\providecommand \@href[1]{\@@startlink{#1}\@@href}%
	\providecommand \@@href[1]{\endgroup#1\@@endlink}%
	\providecommand \@sanitize@url [0]{\catcode `\\12\catcode `\$12\catcode
		`\&12\catcode `\#12\catcode `\^12\catcode `\_12\catcode `\%12\relax}%
	\providecommand \@@startlink[1]{}%
	\providecommand \@@endlink[0]{}%
	\providecommand \url  [0]{\begingroup\@sanitize@url \@url }%
	\providecommand \@url [1]{\endgroup\@href {#1}{\urlprefix }}%
	\providecommand \urlprefix  [0]{URL }%
	\providecommand \Eprint [0]{\href }%
	\providecommand \doibase [0]{https://doi.org/}%
	\providecommand \selectlanguage [0]{\@gobble}%
	\providecommand \bibinfo  [0]{\@secondoftwo}%
	\providecommand \bibfield  [0]{\@secondoftwo}%
	\providecommand \translation [1]{[#1]}%
	\providecommand \BibitemOpen [0]{}%
	\providecommand \bibitemStop [0]{}%
	\providecommand \bibitemNoStop [0]{.\EOS\space}%
	\providecommand \EOS [0]{\spacefactor3000\relax}%
	\providecommand \BibitemShut  [1]{\csname bibitem#1\endcsname}%
	\let\auto@bib@innerbib\@empty
	\bibitem [{\citenamefont {Stewart}(2017)}]{Stewart2017}%
	\BibitemOpen
	\bibfield  {author} {\bibinfo {author} {\bibfnamefont {G.~R.}\ \bibnamefont
			{Stewart}},\ }\href {https://doi.org/10.1080/00018732.2017.1331615}
	{\bibfield  {journal} {\bibinfo  {journal} {Advances in Physics}\ }\textbf
		{\bibinfo {volume} {66}},\ \bibinfo {pages} {75} (\bibinfo {year}
		{2017})}\BibitemShut {NoStop}%
	\bibitem [{\citenamefont {Jiang}\ \emph {et~al.}(2022)\citenamefont {Jiang},
		\citenamefont {Wu}, \citenamefont {Yin}, \citenamefont {Wang}, \citenamefont
		{Hasan}, \citenamefont {Wilson}, \citenamefont {Chen},\ and\ \citenamefont
		{Hu}}]{Jiang2022}%
	\BibitemOpen
	\bibfield  {author} {\bibinfo {author} {\bibfnamefont {K.}~\bibnamefont
			{Jiang}}, \bibinfo {author} {\bibfnamefont {T.}~\bibnamefont {Wu}}, \bibinfo
		{author} {\bibfnamefont {J.-X.}\ \bibnamefont {Yin}}, \bibinfo {author}
		{\bibfnamefont {Z.}~\bibnamefont {Wang}}, \bibinfo {author} {\bibfnamefont
			{M.~Z.}\ \bibnamefont {Hasan}}, \bibinfo {author} {\bibfnamefont {S.~D.}\
			\bibnamefont {Wilson}}, \bibinfo {author} {\bibfnamefont {X.}~\bibnamefont
			{Chen}},\ and\ \bibinfo {author} {\bibfnamefont {J.}~\bibnamefont {Hu}},\
	}\href {https://doi.org/10.1093/nsr/nwac199} {\bibfield  {journal} {\bibinfo
			{journal} {National Science Review}\ }\textbf {\bibinfo {volume} {10}},\
		\bibinfo {pages} {nwac199} (\bibinfo {year} {2022})}\BibitemShut {NoStop}%
	\bibitem [{\citenamefont {Leggett}(1975)}]{Leggett1975}%
	\BibitemOpen
	\bibfield  {author} {\bibinfo {author} {\bibfnamefont {A.~J.}\ \bibnamefont
			{Leggett}},\ }\href {https://doi.org/10.1103/RevModPhys.47.331} {\bibfield
		{journal} {\bibinfo  {journal} {Rev. Mod. Phys.}\ }\textbf {\bibinfo {volume}
			{47}},\ \bibinfo {pages} {331} (\bibinfo {year} {1975})}\BibitemShut
	{NoStop}%
	\bibitem [{\citenamefont {Fisher}\ \emph {et~al.}(1989)\citenamefont {Fisher},
		\citenamefont {Kim}, \citenamefont {Woodfield}, \citenamefont {Phillips},
		\citenamefont {Taillefer}, \citenamefont {Hasselbach}, \citenamefont
		{Flouquet}, \citenamefont {Giorgi},\ and\ \citenamefont
		{Smith}}]{Fisher1989}%
	\BibitemOpen
	\bibfield  {author} {\bibinfo {author} {\bibfnamefont {R.~A.}\ \bibnamefont
			{Fisher}}, \bibinfo {author} {\bibfnamefont {S.}~\bibnamefont {Kim}},
		\bibinfo {author} {\bibfnamefont {B.~F.}\ \bibnamefont {Woodfield}}, \bibinfo
		{author} {\bibfnamefont {N.~E.}\ \bibnamefont {Phillips}}, \bibinfo {author}
		{\bibfnamefont {L.}~\bibnamefont {Taillefer}}, \bibinfo {author}
		{\bibfnamefont {K.}~\bibnamefont {Hasselbach}}, \bibinfo {author}
		{\bibfnamefont {J.}~\bibnamefont {Flouquet}}, \bibinfo {author}
		{\bibfnamefont {A.~L.}\ \bibnamefont {Giorgi}},\ and\ \bibinfo {author}
		{\bibfnamefont {J.~L.}\ \bibnamefont {Smith}},\ }\href
	{https://doi.org/10.1103/PhysRevLett.62.1411} {\bibfield  {journal} {\bibinfo
			{journal} {Phys. Rev. Lett.}\ }\textbf {\bibinfo {volume} {62}},\ \bibinfo
		{pages} {1411} (\bibinfo {year} {1989})}\BibitemShut {NoStop}%
	\bibitem [{\citenamefont {Hasselbach}\ \emph {et~al.}(1989)\citenamefont
		{Hasselbach}, \citenamefont {Taillefer},\ and\ \citenamefont
		{Flouquet}}]{Hasselbach1989}%
	\BibitemOpen
	\bibfield  {author} {\bibinfo {author} {\bibfnamefont {K.}~\bibnamefont
			{Hasselbach}}, \bibinfo {author} {\bibfnamefont {L.}~\bibnamefont
			{Taillefer}},\ and\ \bibinfo {author} {\bibfnamefont {J.}~\bibnamefont
			{Flouquet}},\ }\href {https://doi.org/10.1103/PhysRevLett.63.93} {\bibfield
		{journal} {\bibinfo  {journal} {Phys. Rev. Lett.}\ }\textbf {\bibinfo
			{volume} {63}},\ \bibinfo {pages} {93} (\bibinfo {year} {1989})}\BibitemShut
	{NoStop}%
	\bibitem [{\citenamefont {Adenwalla}\ \emph {et~al.}(1990)\citenamefont
		{Adenwalla}, \citenamefont {Lin}, \citenamefont {Ran}, \citenamefont {Zhao},
		\citenamefont {Ketterson}, \citenamefont {Sauls}, \citenamefont {Taillefer},
		\citenamefont {Hinks}, \citenamefont {Levy},\ and\ \citenamefont
		{Sarma}}]{Adenwalla1990}%
	\BibitemOpen
	\bibfield  {author} {\bibinfo {author} {\bibfnamefont {S.}~\bibnamefont
			{Adenwalla}}, \bibinfo {author} {\bibfnamefont {S.~W.}\ \bibnamefont {Lin}},
		\bibinfo {author} {\bibfnamefont {Q.~Z.}\ \bibnamefont {Ran}}, \bibinfo
		{author} {\bibfnamefont {Z.}~\bibnamefont {Zhao}}, \bibinfo {author}
		{\bibfnamefont {J.~B.}\ \bibnamefont {Ketterson}}, \bibinfo {author}
		{\bibfnamefont {J.~A.}\ \bibnamefont {Sauls}}, \bibinfo {author}
		{\bibfnamefont {L.}~\bibnamefont {Taillefer}}, \bibinfo {author}
		{\bibfnamefont {D.~G.}\ \bibnamefont {Hinks}}, \bibinfo {author}
		{\bibfnamefont {M.}~\bibnamefont {Levy}},\ and\ \bibinfo {author}
		{\bibfnamefont {B.~K.}\ \bibnamefont {Sarma}},\ }\href
	{https://doi.org/10.1103/PhysRevLett.65.2298} {\bibfield  {journal} {\bibinfo
			{journal} {Phys. Rev. Lett.}\ }\textbf {\bibinfo {volume} {65}},\ \bibinfo
		{pages} {2298} (\bibinfo {year} {1990})}\BibitemShut {NoStop}%
	\bibitem [{\citenamefont {Ott}\ \emph {et~al.}(1985)\citenamefont {Ott},
		\citenamefont {Rudigier}, \citenamefont {Fisk},\ and\ \citenamefont
		{Smith}}]{Ott1985}%
	\BibitemOpen
	\bibfield  {author} {\bibinfo {author} {\bibfnamefont {H.~R.}\ \bibnamefont
			{Ott}}, \bibinfo {author} {\bibfnamefont {H.}~\bibnamefont {Rudigier}},
		\bibinfo {author} {\bibfnamefont {Z.}~\bibnamefont {Fisk}},\ and\ \bibinfo
		{author} {\bibfnamefont {J.~L.}\ \bibnamefont {Smith}},\ }\href
	{https://doi.org/10.1103/PhysRevB.31.1651} {\bibfield  {journal} {\bibinfo
			{journal} {Phys. Rev. B}\ }\textbf {\bibinfo {volume} {31}},\ \bibinfo
		{pages} {1651} (\bibinfo {year} {1985})}\BibitemShut {NoStop}%
	\bibitem [{\citenamefont {Khim}\ \emph {et~al.}(2021)\citenamefont {Khim},
		\citenamefont {Landaeta}, \citenamefont {Banda}, \citenamefont {Bannor},
		\citenamefont {Brando}, \citenamefont {Brydon}, \citenamefont {Hafner},
		\citenamefont {Kuechler}, \citenamefont {Cardoso-Gil}, \citenamefont
		{Stockert}, \citenamefont {Mackenzie}, \citenamefont {Agterberg},
		\citenamefont {Geibel},\ and\ \citenamefont {Hassinger}}]{Khim2021}%
	\BibitemOpen
	\bibfield  {author} {\bibinfo {author} {\bibfnamefont {S.}~\bibnamefont
			{Khim}}, \bibinfo {author} {\bibfnamefont {J.~F.}\ \bibnamefont {Landaeta}},
		\bibinfo {author} {\bibfnamefont {J.}~\bibnamefont {Banda}}, \bibinfo
		{author} {\bibfnamefont {N.}~\bibnamefont {Bannor}}, \bibinfo {author}
		{\bibfnamefont {M.}~\bibnamefont {Brando}}, \bibinfo {author} {\bibfnamefont
			{P.~M.~R.}\ \bibnamefont {Brydon}}, \bibinfo {author} {\bibfnamefont
			{D.}~\bibnamefont {Hafner}}, \bibinfo {author} {\bibfnamefont
			{R.}~\bibnamefont {Kuechler}}, \bibinfo {author} {\bibfnamefont
			{R.}~\bibnamefont {Cardoso-Gil}}, \bibinfo {author} {\bibfnamefont
			{U.}~\bibnamefont {Stockert}}, \bibinfo {author} {\bibfnamefont {A.~P.}\
			\bibnamefont {Mackenzie}}, \bibinfo {author} {\bibfnamefont {D.~F.}\
			\bibnamefont {Agterberg}}, \bibinfo {author} {\bibfnamefont {C.}~\bibnamefont
			{Geibel}},\ and\ \bibinfo {author} {\bibfnamefont {E.}~\bibnamefont
			{Hassinger}},\ }\href {https://doi.org/{10.1126/science.abe7518}} {\bibfield
		{journal} {\bibinfo  {journal} {Science}\ }\textbf {\bibinfo {volume}
			{{373}}},\ \bibinfo {pages} {{1012+}} (\bibinfo {year} {{2021}})}\BibitemShut
	{NoStop}%
	\bibitem [{\citenamefont {Ran}\ \emph {et~al.}(2019{\natexlab{a}})\citenamefont
		{Ran}, \citenamefont {Eckberg}, \citenamefont {Ding}, \citenamefont
		{Furukawa}, \citenamefont {Metz}, \citenamefont {Saha}, \citenamefont {Liu},
		\citenamefont {Zic}, \citenamefont {Kim}, \citenamefont {Paglione},\ and\
		\citenamefont {Butch}}]{Ran2019}%
	\BibitemOpen
	\bibfield  {author} {\bibinfo {author} {\bibfnamefont {S.}~\bibnamefont
			{Ran}}, \bibinfo {author} {\bibfnamefont {C.}~\bibnamefont {Eckberg}},
		\bibinfo {author} {\bibfnamefont {Q.-P.}\ \bibnamefont {Ding}}, \bibinfo
		{author} {\bibfnamefont {Y.}~\bibnamefont {Furukawa}}, \bibinfo {author}
		{\bibfnamefont {T.}~\bibnamefont {Metz}}, \bibinfo {author} {\bibfnamefont
			{S.~R.}\ \bibnamefont {Saha}}, \bibinfo {author} {\bibfnamefont {I.-L.}\
			\bibnamefont {Liu}}, \bibinfo {author} {\bibfnamefont {M.}~\bibnamefont
			{Zic}}, \bibinfo {author} {\bibfnamefont {H.}~\bibnamefont {Kim}}, \bibinfo
		{author} {\bibfnamefont {J.}~\bibnamefont {Paglione}},\ and\ \bibinfo
		{author} {\bibfnamefont {N.~P.}\ \bibnamefont {Butch}},\ }\href
	{https://doi.org/10.1126/science.aav8645} {\bibfield  {journal} {\bibinfo
			{journal} {Science}\ }\textbf {\bibinfo {volume} {365}},\ \bibinfo {pages}
		{684} (\bibinfo {year} {2019}{\natexlab{a}})}\BibitemShut {NoStop}%
	\bibitem [{\citenamefont {Aoki}\ \emph {et~al.}(2019)\citenamefont {Aoki},
		\citenamefont {Nakamura}, \citenamefont {Honda}, \citenamefont {Li},
		\citenamefont {Homma}, \citenamefont {Shimizu}, \citenamefont {Sato},
		\citenamefont {Knebel}, \citenamefont {Brison}, \citenamefont {Pourret},
		\citenamefont {Braithwaite}, \citenamefont {Lapertot}, \citenamefont {Niu},
		\citenamefont {Vali\v{s}ka}, \citenamefont {Harima},\ and\ \citenamefont
		{Flouquet}}]{Aoki2019}%
	\BibitemOpen
	\bibfield  {author} {\bibinfo {author} {\bibfnamefont {D.}~\bibnamefont
			{Aoki}}, \bibinfo {author} {\bibfnamefont {A.}~\bibnamefont {Nakamura}},
		\bibinfo {author} {\bibfnamefont {F.}~\bibnamefont {Honda}}, \bibinfo
		{author} {\bibfnamefont {D.}~\bibnamefont {Li}}, \bibinfo {author}
		{\bibfnamefont {Y.}~\bibnamefont {Homma}}, \bibinfo {author} {\bibfnamefont
			{Y.}~\bibnamefont {Shimizu}}, \bibinfo {author} {\bibfnamefont {Y.~J.}\
			\bibnamefont {Sato}}, \bibinfo {author} {\bibfnamefont {G.}~\bibnamefont
			{Knebel}}, \bibinfo {author} {\bibfnamefont {J.-P.}\ \bibnamefont {Brison}},
		\bibinfo {author} {\bibfnamefont {A.}~\bibnamefont {Pourret}}, \bibinfo
		{author} {\bibfnamefont {D.}~\bibnamefont {Braithwaite}}, \bibinfo {author}
		{\bibfnamefont {G.}~\bibnamefont {Lapertot}}, \bibinfo {author}
		{\bibfnamefont {Q.}~\bibnamefont {Niu}}, \bibinfo {author} {\bibfnamefont
			{M.}~\bibnamefont {Vali\v{s}ka}}, \bibinfo {author} {\bibfnamefont
			{H.}~\bibnamefont {Harima}},\ and\ \bibinfo {author} {\bibfnamefont
			{J.}~\bibnamefont {Flouquet}},\ }\href
	{https://doi.org/10.7566/JPSJ.88.043702} {\bibfield  {journal} {\bibinfo
			{journal} {J. Phys. Soc. Jpn.}\ }\textbf {\bibinfo {volume} {88}},\ \bibinfo
		{pages} {043702} (\bibinfo {year} {2019})}\BibitemShut {NoStop}%
	\bibitem [{\citenamefont {Aoki}\ \emph {et~al.}(2022)\citenamefont {Aoki},
		\citenamefont {Brison}, \citenamefont {Flouquet}, \citenamefont {Ishida},
		\citenamefont {Knebel}, \citenamefont {Tokunaga},\ and\ \citenamefont
		{Yanase}}]{Aoki2022review}%
	\BibitemOpen
	\bibfield  {author} {\bibinfo {author} {\bibfnamefont {D.}~\bibnamefont
			{Aoki}}, \bibinfo {author} {\bibfnamefont {J.-P.}\ \bibnamefont {Brison}},
		\bibinfo {author} {\bibfnamefont {J.}~\bibnamefont {Flouquet}}, \bibinfo
		{author} {\bibfnamefont {K.}~\bibnamefont {Ishida}}, \bibinfo {author}
		{\bibfnamefont {G.}~\bibnamefont {Knebel}}, \bibinfo {author} {\bibfnamefont
			{Y.}~\bibnamefont {Tokunaga}},\ and\ \bibinfo {author} {\bibfnamefont
			{Y.}~\bibnamefont {Yanase}},\ }\href
	{https://doi.org/10.1088/1361-648x/ac5863} {\bibfield  {journal} {\bibinfo
			{journal} {J. Phys.: Condens. Matter}\ }\textbf {\bibinfo {volume} {34}},\
		\bibinfo {pages} {243002} (\bibinfo {year} {2022})}\BibitemShut {NoStop}%
	\bibitem [{\citenamefont {Lewin}\ \emph {et~al.}(2023)\citenamefont {Lewin},
		\citenamefont {Frank}, \citenamefont {Ran}, \citenamefont {Paglione},\ and\
		\citenamefont {Butch}}]{Lewin2023}%
	\BibitemOpen
	\bibfield  {author} {\bibinfo {author} {\bibfnamefont {S.~K.}\ \bibnamefont
			{Lewin}}, \bibinfo {author} {\bibfnamefont {C.~E.}\ \bibnamefont {Frank}},
		\bibinfo {author} {\bibfnamefont {S.}~\bibnamefont {Ran}}, \bibinfo {author}
		{\bibfnamefont {J.}~\bibnamefont {Paglione}},\ and\ \bibinfo {author}
		{\bibfnamefont {N.~P.}\ \bibnamefont {Butch}},\ }\href
	{https://doi.org/10.1088/1361-6633/acfb93} {\bibfield  {journal} {\bibinfo
			{journal} {Rep. Progr. Phys.}\ }\textbf {\bibinfo {volume} {86}},\ \bibinfo
		{pages} {114501} (\bibinfo {year} {2023})}\BibitemShut {NoStop}%
	\bibitem [{\citenamefont {Braithwaite}\ \emph {et~al.}(2019)\citenamefont
		{Braithwaite}, \citenamefont {Vali{\v{s}}ka}, \citenamefont {Knebel},
		\citenamefont {Lapertot}, \citenamefont {Brison}, \citenamefont {Pourret},
		\citenamefont {Zhitomirsky}, \citenamefont {Flouquet}, \citenamefont
		{Honda},\ and\ \citenamefont {Aoki}}]{Braithwaite2019}%
	\BibitemOpen
	\bibfield  {author} {\bibinfo {author} {\bibfnamefont {D.}~\bibnamefont
			{Braithwaite}}, \bibinfo {author} {\bibfnamefont {M.}~\bibnamefont
			{Vali{\v{s}}ka}}, \bibinfo {author} {\bibfnamefont {G.}~\bibnamefont
			{Knebel}}, \bibinfo {author} {\bibfnamefont {G.}~\bibnamefont {Lapertot}},
		\bibinfo {author} {\bibfnamefont {J.~P.}\ \bibnamefont {Brison}}, \bibinfo
		{author} {\bibfnamefont {A.}~\bibnamefont {Pourret}}, \bibinfo {author}
		{\bibfnamefont {M.~E.}\ \bibnamefont {Zhitomirsky}}, \bibinfo {author}
		{\bibfnamefont {J.}~\bibnamefont {Flouquet}}, \bibinfo {author}
		{\bibfnamefont {F.}~\bibnamefont {Honda}},\ and\ \bibinfo {author}
		{\bibfnamefont {D.}~\bibnamefont {Aoki}},\ }\bibfield  {journal} {\bibinfo
		{journal} {Commun Phys}\ }\textbf {\bibinfo {volume} {2}},\ \href
	{https://doi.org/10.1038/s42005-019-0248-z} {10.1038/s42005-019-0248-z}
	(\bibinfo {year} {2019})\BibitemShut {NoStop}%
	\bibitem [{\citenamefont {Thomas}\ \emph {et~al.}(2024)\citenamefont {Thomas},
		\citenamefont {Santos}, \citenamefont {Christensen}, \citenamefont {Asaba},
		\citenamefont {Ronning}, \citenamefont {Thompson}, \citenamefont {Bauer},
		\citenamefont {Fernandes}, \citenamefont {Fabbris},\ and\ \citenamefont
		{Rosa}}]{Thomas2020}%
	\BibitemOpen
	\bibfield  {author} {\bibinfo {author} {\bibfnamefont {S.~M.}\ \bibnamefont
			{Thomas}}, \bibinfo {author} {\bibfnamefont {F.~B.}\ \bibnamefont {Santos}},
		\bibinfo {author} {\bibfnamefont {M.~H.}\ \bibnamefont {Christensen}},
		\bibinfo {author} {\bibfnamefont {T.}~\bibnamefont {Asaba}}, \bibinfo
		{author} {\bibfnamefont {F.}~\bibnamefont {Ronning}}, \bibinfo {author}
		{\bibfnamefont {J.~D.}\ \bibnamefont {Thompson}}, \bibinfo {author}
		{\bibfnamefont {E.~D.}\ \bibnamefont {Bauer}}, \bibinfo {author}
		{\bibfnamefont {R.~M.}\ \bibnamefont {Fernandes}}, \bibinfo {author}
		{\bibfnamefont {G.}~\bibnamefont {Fabbris}},\ and\ \bibinfo {author}
		{\bibfnamefont {P.~F.~S.}\ \bibnamefont {Rosa}},\ }\href
	{https://doi.org/10.1126/sciadv.abc8709} {\bibfield  {journal} {\bibinfo
			{journal} {Sci. Adv.}\ }\textbf {\bibinfo {volume} {6}},\ \bibinfo {pages}
		{eabc8709} (\bibinfo {year} {2024})}\BibitemShut {NoStop}%
	\bibitem [{\citenamefont {Ran}\ \emph {et~al.}(2019{\natexlab{b}})\citenamefont
		{Ran}, \citenamefont {Liu}, \citenamefont {Eo}, \citenamefont {Campbell},
		\citenamefont {Neves}, \citenamefont {Fuhrman}, \citenamefont {Saha},
		\citenamefont {Eckberg}, \citenamefont {Kim}, \citenamefont {Graf},
		\citenamefont {Balakirev}, \citenamefont {Singleton}, \citenamefont
		{Paglione},\ and\ \citenamefont {Butch}}]{Ran2019a}%
	\BibitemOpen
	\bibfield  {author} {\bibinfo {author} {\bibfnamefont {S.}~\bibnamefont
			{Ran}}, \bibinfo {author} {\bibfnamefont {I.-L.}\ \bibnamefont {Liu}},
		\bibinfo {author} {\bibfnamefont {Y.~S.}\ \bibnamefont {Eo}}, \bibinfo
		{author} {\bibfnamefont {D.~J.}\ \bibnamefont {Campbell}}, \bibinfo {author}
		{\bibfnamefont {P.~M.}\ \bibnamefont {Neves}}, \bibinfo {author}
		{\bibfnamefont {W.~T.}\ \bibnamefont {Fuhrman}}, \bibinfo {author}
		{\bibfnamefont {S.~R.}\ \bibnamefont {Saha}}, \bibinfo {author}
		{\bibfnamefont {C.}~\bibnamefont {Eckberg}}, \bibinfo {author} {\bibfnamefont
			{H.}~\bibnamefont {Kim}}, \bibinfo {author} {\bibfnamefont {D.}~\bibnamefont
			{Graf}}, \bibinfo {author} {\bibfnamefont {F.}~\bibnamefont {Balakirev}},
		\bibinfo {author} {\bibfnamefont {J.}~\bibnamefont {Singleton}}, \bibinfo
		{author} {\bibfnamefont {J.}~\bibnamefont {Paglione}},\ and\ \bibinfo
		{author} {\bibfnamefont {N.~P.}\ \bibnamefont {Butch}},\ }\href
	{https://doi.org/10.1038/s41567-019-0670-x} {\bibfield  {journal} {\bibinfo
			{journal} {Nat. Phys.}\ }\textbf {\bibinfo {volume} {15}},\ \bibinfo {pages}
		{1250} (\bibinfo {year} {2019}{\natexlab{b}})}\BibitemShut {NoStop}%
	\bibitem [{\citenamefont {Knebel}\ \emph {et~al.}(2019)\citenamefont {Knebel},
		\citenamefont {Knafo}, \citenamefont {Pourret}, \citenamefont {Niu},
		\citenamefont {Vališka}, \citenamefont {Braithwaite}, \citenamefont
		{Lapertot}, \citenamefont {Nardone}, \citenamefont {Zitouni}, \citenamefont
		{Mishra}, \citenamefont {Sheikin}, \citenamefont {Seyfarth}, \citenamefont
		{Brison}, \citenamefont {Aoki},\ and\ \citenamefont {Flouquet}}]{Knebel2019}%
	\BibitemOpen
	\bibfield  {author} {\bibinfo {author} {\bibfnamefont {G.}~\bibnamefont
			{Knebel}}, \bibinfo {author} {\bibfnamefont {W.}~\bibnamefont {Knafo}},
		\bibinfo {author} {\bibfnamefont {A.}~\bibnamefont {Pourret}}, \bibinfo
		{author} {\bibfnamefont {Q.}~\bibnamefont {Niu}}, \bibinfo {author}
		{\bibfnamefont {M.}~\bibnamefont {Vališka}}, \bibinfo {author}
		{\bibfnamefont {D.}~\bibnamefont {Braithwaite}}, \bibinfo {author}
		{\bibfnamefont {G.}~\bibnamefont {Lapertot}}, \bibinfo {author}
		{\bibfnamefont {M.}~\bibnamefont {Nardone}}, \bibinfo {author} {\bibfnamefont
			{A.}~\bibnamefont {Zitouni}}, \bibinfo {author} {\bibfnamefont
			{S.}~\bibnamefont {Mishra}}, \bibinfo {author} {\bibfnamefont
			{I.}~\bibnamefont {Sheikin}}, \bibinfo {author} {\bibfnamefont
			{G.}~\bibnamefont {Seyfarth}}, \bibinfo {author} {\bibfnamefont {J.-P.}\
			\bibnamefont {Brison}}, \bibinfo {author} {\bibfnamefont {D.}~\bibnamefont
			{Aoki}},\ and\ \bibinfo {author} {\bibfnamefont {J.}~\bibnamefont
			{Flouquet}},\ }\href {https://doi.org/10.7566/JPSJ.88.063707} {\bibfield
		{journal} {\bibinfo  {journal} {J. Phys. Soc. Jpn.}\ }\textbf {\bibinfo
			{volume} {88}},\ \bibinfo {pages} {063707} (\bibinfo {year}
		{2019})}\BibitemShut {NoStop}%
	\bibitem [{\citenamefont {Knafo}\ \emph {et~al.}(2019)\citenamefont {Knafo},
		\citenamefont {Vališka}, \citenamefont {Braithwaite}, \citenamefont
		{Lapertot}, \citenamefont {Knebel}, \citenamefont {Pourret}, \citenamefont
		{Brison}, \citenamefont {Flouquet},\ and\ \citenamefont {Aoki}}]{Knafo2019}%
	\BibitemOpen
	\bibfield  {author} {\bibinfo {author} {\bibfnamefont {W.}~\bibnamefont
			{Knafo}}, \bibinfo {author} {\bibfnamefont {M.}~\bibnamefont {Vališka}},
		\bibinfo {author} {\bibfnamefont {D.}~\bibnamefont {Braithwaite}}, \bibinfo
		{author} {\bibfnamefont {G.}~\bibnamefont {Lapertot}}, \bibinfo {author}
		{\bibfnamefont {G.}~\bibnamefont {Knebel}}, \bibinfo {author} {\bibfnamefont
			{A.}~\bibnamefont {Pourret}}, \bibinfo {author} {\bibfnamefont {J.-P.}\
			\bibnamefont {Brison}}, \bibinfo {author} {\bibfnamefont {J.}~\bibnamefont
			{Flouquet}},\ and\ \bibinfo {author} {\bibfnamefont {D.}~\bibnamefont
			{Aoki}},\ }\href {https://doi.org/10.7566/JPSJ.88.063705} {\bibfield
		{journal} {\bibinfo  {journal} {J. Phys. Soc. Jpn.}\ }\textbf {\bibinfo
			{volume} {88}},\ \bibinfo {pages} {063705} (\bibinfo {year}
		{2019})}\BibitemShut {NoStop}%
	\bibitem [{\citenamefont {Rosuel}\ \emph {et~al.}(2023)\citenamefont {Rosuel},
		\citenamefont {Marcenat}, \citenamefont {Knebel}, \citenamefont {Klein},
		\citenamefont {Pourret}, \citenamefont {Marquardt}, \citenamefont {Niu},
		\citenamefont {Rousseau}, \citenamefont {Demuer}, \citenamefont {Seyfarth},
		\citenamefont {Lapertot}, \citenamefont {Aoki}, \citenamefont {Braithwaite},
		\citenamefont {Flouquet},\ and\ \citenamefont {Brison}}]{Rosuel2023}%
	\BibitemOpen
	\bibfield  {author} {\bibinfo {author} {\bibfnamefont {A.}~\bibnamefont
			{Rosuel}}, \bibinfo {author} {\bibfnamefont {C.}~\bibnamefont {Marcenat}},
		\bibinfo {author} {\bibfnamefont {G.}~\bibnamefont {Knebel}}, \bibinfo
		{author} {\bibfnamefont {T.}~\bibnamefont {Klein}}, \bibinfo {author}
		{\bibfnamefont {A.}~\bibnamefont {Pourret}}, \bibinfo {author} {\bibfnamefont
			{N.}~\bibnamefont {Marquardt}}, \bibinfo {author} {\bibfnamefont
			{Q.}~\bibnamefont {Niu}}, \bibinfo {author} {\bibfnamefont {S.}~\bibnamefont
			{Rousseau}}, \bibinfo {author} {\bibfnamefont {A.}~\bibnamefont {Demuer}},
		\bibinfo {author} {\bibfnamefont {G.}~\bibnamefont {Seyfarth}}, \bibinfo
		{author} {\bibfnamefont {G.}~\bibnamefont {Lapertot}}, \bibinfo {author}
		{\bibfnamefont {D.}~\bibnamefont {Aoki}}, \bibinfo {author} {\bibfnamefont
			{D.}~\bibnamefont {Braithwaite}}, \bibinfo {author} {\bibfnamefont
			{J.}~\bibnamefont {Flouquet}},\ and\ \bibinfo {author} {\bibfnamefont
			{J.~P.}\ \bibnamefont {Brison}},\ }\href
	{https://doi.org/10.1103/PhysRevX.13.011022} {\bibfield  {journal} {\bibinfo
			{journal} {Phys. Rev. X}\ }\textbf {\bibinfo {volume} {13}},\ \bibinfo
		{pages} {011022} (\bibinfo {year} {2023})}\BibitemShut {NoStop}%
	\bibitem [{\citenamefont {Sauls}(1994)}]{Sauls1994}%
	\BibitemOpen
	\bibfield  {author} {\bibinfo {author} {\bibfnamefont {J.}~\bibnamefont
			{Sauls}},\ }\href {https://doi.org/10.1080/00018739400101475} {\bibfield
		{journal} {\bibinfo  {journal} {Adv. Phys.}\ }\textbf {\bibinfo {volume}
			{43}},\ \bibinfo {pages} {113} (\bibinfo {year} {1994})},\ \Eprint
	{https://arxiv.org/abs/http://dx.doi.org/10.1080/00018739400101475}
	{http://dx.doi.org/10.1080/00018739400101475} \BibitemShut {NoStop}%
	\bibitem [{\citenamefont {Tokunaga}\ \emph {et~al.}(2023)\citenamefont
		{Tokunaga}, \citenamefont {Sakai}, \citenamefont {Kambe}, \citenamefont
		{Opletal}, \citenamefont {Tokiwa}, \citenamefont {Haga}, \citenamefont
		{Kitagawa}, \citenamefont {Ishida}, \citenamefont {Aoki}, \citenamefont
		{Knebel}, \citenamefont {Lapertot}, \citenamefont {Kr\"amer},\ and\
		\citenamefont {Horvati\ifmmode~\acute{c}\else \'{c}\fi{}}}]{Tokunaga2023}%
	\BibitemOpen
	\bibfield  {author} {\bibinfo {author} {\bibfnamefont {Y.}~\bibnamefont
			{Tokunaga}}, \bibinfo {author} {\bibfnamefont {H.}~\bibnamefont {Sakai}},
		\bibinfo {author} {\bibfnamefont {S.}~\bibnamefont {Kambe}}, \bibinfo
		{author} {\bibfnamefont {P.}~\bibnamefont {Opletal}}, \bibinfo {author}
		{\bibfnamefont {Y.}~\bibnamefont {Tokiwa}}, \bibinfo {author} {\bibfnamefont
			{Y.}~\bibnamefont {Haga}}, \bibinfo {author} {\bibfnamefont {S.}~\bibnamefont
			{Kitagawa}}, \bibinfo {author} {\bibfnamefont {K.}~\bibnamefont {Ishida}},
		\bibinfo {author} {\bibfnamefont {D.}~\bibnamefont {Aoki}}, \bibinfo {author}
		{\bibfnamefont {G.}~\bibnamefont {Knebel}}, \bibinfo {author} {\bibfnamefont
			{G.}~\bibnamefont {Lapertot}}, \bibinfo {author} {\bibfnamefont
			{S.}~\bibnamefont {Kr\"amer}},\ and\ \bibinfo {author} {\bibfnamefont
			{M.}~\bibnamefont {Horvati\ifmmode~\acute{c}\else \'{c}\fi{}}},\ }\href
	{https://doi.org/10.1103/PhysRevLett.131.226503} {\bibfield  {journal}
		{\bibinfo  {journal} {Phys. Rev. Lett.}\ }\textbf {\bibinfo {volume} {131}},\
		\bibinfo {pages} {226503} (\bibinfo {year} {2023})}\BibitemShut {NoStop}%
	\bibitem [{\citenamefont {Kinjo}\ \emph
		{et~al.}(2023{\natexlab{a}})\citenamefont {Kinjo}, \citenamefont
		{Fujibayashi}, \citenamefont {Kitagawa}, \citenamefont {Ishida},
		\citenamefont {Tokunaga}, \citenamefont {Sakai}, \citenamefont {Kambe},
		\citenamefont {Nakamura}, \citenamefont {Shimizu}, \citenamefont {Homma},
		\citenamefont {Li}, \citenamefont {Honda}, \citenamefont {Aoki},
		\citenamefont {Hiraki}, \citenamefont {Kimata},\ and\ \citenamefont
		{Sasaki}}]{Kinjo2023PRB}%
	\BibitemOpen
	\bibfield  {author} {\bibinfo {author} {\bibfnamefont {K.}~\bibnamefont
			{Kinjo}}, \bibinfo {author} {\bibfnamefont {H.}~\bibnamefont {Fujibayashi}},
		\bibinfo {author} {\bibfnamefont {S.}~\bibnamefont {Kitagawa}}, \bibinfo
		{author} {\bibfnamefont {K.}~\bibnamefont {Ishida}}, \bibinfo {author}
		{\bibfnamefont {Y.}~\bibnamefont {Tokunaga}}, \bibinfo {author}
		{\bibfnamefont {H.}~\bibnamefont {Sakai}}, \bibinfo {author} {\bibfnamefont
			{S.}~\bibnamefont {Kambe}}, \bibinfo {author} {\bibfnamefont
			{A.}~\bibnamefont {Nakamura}}, \bibinfo {author} {\bibfnamefont
			{Y.}~\bibnamefont {Shimizu}}, \bibinfo {author} {\bibfnamefont
			{Y.}~\bibnamefont {Homma}}, \bibinfo {author} {\bibfnamefont {D.~X.}\
			\bibnamefont {Li}}, \bibinfo {author} {\bibfnamefont {F.}~\bibnamefont
			{Honda}}, \bibinfo {author} {\bibfnamefont {D.}~\bibnamefont {Aoki}},
		\bibinfo {author} {\bibfnamefont {K.}~\bibnamefont {Hiraki}}, \bibinfo
		{author} {\bibfnamefont {M.}~\bibnamefont {Kimata}},\ and\ \bibinfo {author}
		{\bibfnamefont {T.}~\bibnamefont {Sasaki}},\ }\href
	{https://doi.org/10.1103/PhysRevB.107.L060502} {\bibfield  {journal}
		{\bibinfo  {journal} {Phys. Rev. B}\ }\textbf {\bibinfo {volume} {107}},\
		\bibinfo {pages} {L060502} (\bibinfo {year}
		{2023}{\natexlab{a}})}\BibitemShut {NoStop}%
	\bibitem [{\citenamefont {Sakai}\ \emph {et~al.}(2023)\citenamefont {Sakai},
		\citenamefont {Tokiwa}, \citenamefont {Opletal}, \citenamefont {Kimata},
		\citenamefont {Awaji}, \citenamefont {Sasaki}, \citenamefont {Aoki},
		\citenamefont {Kambe}, \citenamefont {Tokunaga},\ and\ \citenamefont
		{Haga}}]{Sakai2023}%
	\BibitemOpen
	\bibfield  {author} {\bibinfo {author} {\bibfnamefont {H.}~\bibnamefont
			{Sakai}}, \bibinfo {author} {\bibfnamefont {Y.}~\bibnamefont {Tokiwa}},
		\bibinfo {author} {\bibfnamefont {P.}~\bibnamefont {Opletal}}, \bibinfo
		{author} {\bibfnamefont {M.}~\bibnamefont {Kimata}}, \bibinfo {author}
		{\bibfnamefont {S.}~\bibnamefont {Awaji}}, \bibinfo {author} {\bibfnamefont
			{T.}~\bibnamefont {Sasaki}}, \bibinfo {author} {\bibfnamefont
			{D.}~\bibnamefont {Aoki}}, \bibinfo {author} {\bibfnamefont {S.}~\bibnamefont
			{Kambe}}, \bibinfo {author} {\bibfnamefont {Y.}~\bibnamefont {Tokunaga}},\
		and\ \bibinfo {author} {\bibfnamefont {Y.}~\bibnamefont {Haga}},\ }\href
	{https://doi.org/10.1103/PhysRevLett.130.196002} {\bibfield  {journal}
		{\bibinfo  {journal} {Phys. Rev. Lett.}\ }\textbf {\bibinfo {volume} {130}},\
		\bibinfo {pages} {196002} (\bibinfo {year} {2023})}\BibitemShut {NoStop}%
	\bibitem [{\citenamefont {Aoki}\ \emph {et~al.}(2020)\citenamefont {Aoki},
		\citenamefont {Honda}, \citenamefont {Knebel}, \citenamefont {Braithwaite},
		\citenamefont {Nakamura}, \citenamefont {Li}, \citenamefont {Homma},
		\citenamefont {Shimizu}, \citenamefont {Sato}, \citenamefont {Brison},\ and\
		\citenamefont {Flouquet}}]{Aoki2020}%
	\BibitemOpen
	\bibfield  {author} {\bibinfo {author} {\bibfnamefont {D.}~\bibnamefont
			{Aoki}}, \bibinfo {author} {\bibfnamefont {F.}~\bibnamefont {Honda}},
		\bibinfo {author} {\bibfnamefont {G.}~\bibnamefont {Knebel}}, \bibinfo
		{author} {\bibfnamefont {D.}~\bibnamefont {Braithwaite}}, \bibinfo {author}
		{\bibfnamefont {A.}~\bibnamefont {Nakamura}}, \bibinfo {author}
		{\bibfnamefont {D.~X.}\ \bibnamefont {Li}}, \bibinfo {author} {\bibfnamefont
			{Y.}~\bibnamefont {Homma}}, \bibinfo {author} {\bibfnamefont
			{Y.}~\bibnamefont {Shimizu}}, \bibinfo {author} {\bibfnamefont {Y.~J.}\
			\bibnamefont {Sato}}, \bibinfo {author} {\bibfnamefont {J.~P.}\ \bibnamefont
			{Brison}},\ and\ \bibinfo {author} {\bibfnamefont {J.}~\bibnamefont
			{Flouquet}},\ }\href {https://doi.org/10.7566/JPSJ.89.053705} {\bibfield
		{journal} {\bibinfo  {journal} {J. Phys. Soc. Jpn.}\ }\textbf {\bibinfo
			{volume} {89}},\ \bibinfo {pages} {1} (\bibinfo {year} {2020})}\BibitemShut
	{NoStop}%
	\bibitem [{\citenamefont {Lin}\ \emph {et~al.}(2020)\citenamefont {Lin},
		\citenamefont {Campbell}, \citenamefont {Ran}, \citenamefont {Liu},
		\citenamefont {Kim}, \citenamefont {Nevidomskyy}, \citenamefont {Graf},
		\citenamefont {Butch},\ and\ \citenamefont {Paglione}}]{Lin2020}%
	\BibitemOpen
	\bibfield  {author} {\bibinfo {author} {\bibfnamefont {W.-C.}\ \bibnamefont
			{Lin}}, \bibinfo {author} {\bibfnamefont {D.~J.}\ \bibnamefont {Campbell}},
		\bibinfo {author} {\bibfnamefont {S.}~\bibnamefont {Ran}}, \bibinfo {author}
		{\bibfnamefont {I.-L.}\ \bibnamefont {Liu}}, \bibinfo {author} {\bibfnamefont
			{H.}~\bibnamefont {Kim}}, \bibinfo {author} {\bibfnamefont {A.~H.}\
			\bibnamefont {Nevidomskyy}}, \bibinfo {author} {\bibfnamefont
			{D.}~\bibnamefont {Graf}}, \bibinfo {author} {\bibfnamefont {N.~P.}\
			\bibnamefont {Butch}},\ and\ \bibinfo {author} {\bibfnamefont
			{J.}~\bibnamefont {Paglione}},\ }\href
	{https://doi.org/10.1038/s41535-020-00270-w} {\bibfield  {journal} {\bibinfo
			{journal} {npj Quantum Materials}\ }\textbf {\bibinfo {volume} {5}},\
		\bibinfo {pages} {68} (\bibinfo {year} {2020})}\BibitemShut {NoStop}%
	\bibitem [{\citenamefont {Landau}\ and\ \citenamefont
		{Lifshitz}(1982)}]{LandauLifshitzVol5}%
	\BibitemOpen
	\bibfield  {author} {\bibinfo {author} {\bibfnamefont {L.~D.}\ \bibnamefont
			{Landau}}\ and\ \bibinfo {author} {\bibfnamefont {E.~M.}\ \bibnamefont
			{Lifshitz}},\ }\href@noop {} {\emph {\bibinfo {title} {Course of Theoretical
				Physics, Statistical Physics Part 1}}},\ Vol.~\bibinfo {volume} {5}\
	(\bibinfo  {publisher} {Pergamon Press},\ \bibinfo {address} {Oxford},\
	\bibinfo {year} {1982})\BibitemShut {NoStop}%
	\bibitem [{\citenamefont {Tokiwa}\ \emph {et~al.}(2023)\citenamefont {Tokiwa},
		\citenamefont {Sakai}, \citenamefont {Kambe}, \citenamefont {Opletal},
		\citenamefont {Yamamoto}, \citenamefont {Kimata}, \citenamefont {Awaji},
		\citenamefont {Sasaki}, \citenamefont {Yanase}, \citenamefont {Haga},\ and\
		\citenamefont {Tokunaga}}]{Tokiwa2023}%
	\BibitemOpen
	\bibfield  {author} {\bibinfo {author} {\bibfnamefont {Y.}~\bibnamefont
			{Tokiwa}}, \bibinfo {author} {\bibfnamefont {H.}~\bibnamefont {Sakai}},
		\bibinfo {author} {\bibfnamefont {S.}~\bibnamefont {Kambe}}, \bibinfo
		{author} {\bibfnamefont {P.}~\bibnamefont {Opletal}}, \bibinfo {author}
		{\bibfnamefont {E.}~\bibnamefont {Yamamoto}}, \bibinfo {author}
		{\bibfnamefont {M.}~\bibnamefont {Kimata}}, \bibinfo {author} {\bibfnamefont
			{S.}~\bibnamefont {Awaji}}, \bibinfo {author} {\bibfnamefont
			{T.}~\bibnamefont {Sasaki}}, \bibinfo {author} {\bibfnamefont
			{Y.}~\bibnamefont {Yanase}}, \bibinfo {author} {\bibfnamefont
			{Y.}~\bibnamefont {Haga}},\ and\ \bibinfo {author} {\bibfnamefont
			{Y.}~\bibnamefont {Tokunaga}},\ }\href
	{https://doi.org/10.1103/PhysRevB.108.144502} {\bibfield  {journal} {\bibinfo
			{journal} {Phys. Rev. B}\ }\textbf {\bibinfo {volume} {108}},\ \bibinfo
		{pages} {144502} (\bibinfo {year} {2023})}\BibitemShut {NoStop}%
	\bibitem [{\citenamefont {Sato}\ and\ \citenamefont {Ando}(2017)}]{Sato2017}%
	\BibitemOpen
	\bibfield  {author} {\bibinfo {author} {\bibfnamefont {M.}~\bibnamefont
			{Sato}}\ and\ \bibinfo {author} {\bibfnamefont {Y.}~\bibnamefont {Ando}},\
	}\href {https://doi.org/10.1088/1361-6633/aa6ac7} {\bibfield  {journal}
		{\bibinfo  {journal} {Reports on Progress in Physics}\ }\textbf {\bibinfo
			{volume} {80}},\ \bibinfo {pages} {076501} (\bibinfo {year} {2017})},\
	\Eprint {https://arxiv.org/abs/1608.03395} {arXiv:1608.03395} \BibitemShut
	{NoStop}%
	\bibitem [{\citenamefont {Bae}\ \emph {et~al.}(2021)\citenamefont {Bae},
		\citenamefont {Kim}, \citenamefont {Eo}, \citenamefont {Ran}, \citenamefont
		{Liu}, \citenamefont {Fuhrman}, \citenamefont {Paglione}, \citenamefont
		{Butch},\ and\ \citenamefont {Anlage}}]{Bae2021}%
	\BibitemOpen
	\bibfield  {author} {\bibinfo {author} {\bibfnamefont {S.}~\bibnamefont
			{Bae}}, \bibinfo {author} {\bibfnamefont {H.}~\bibnamefont {Kim}}, \bibinfo
		{author} {\bibfnamefont {Y.~S.}\ \bibnamefont {Eo}}, \bibinfo {author}
		{\bibfnamefont {S.}~\bibnamefont {Ran}}, \bibinfo {author} {\bibfnamefont
			{I.-l.}\ \bibnamefont {Liu}}, \bibinfo {author} {\bibfnamefont {W.~T.}\
			\bibnamefont {Fuhrman}}, \bibinfo {author} {\bibfnamefont {J.}~\bibnamefont
			{Paglione}}, \bibinfo {author} {\bibfnamefont {N.~P.}\ \bibnamefont
			{Butch}},\ and\ \bibinfo {author} {\bibfnamefont {S.~M.}\ \bibnamefont
			{Anlage}},\ }\href {https://doi.org/10.1038/s41467-021-22906-6} {\bibfield
		{journal} {\bibinfo  {journal} {Nat. Commun.}\ }\textbf {\bibinfo {volume}
			{12}},\ \bibinfo {pages} {2644} (\bibinfo {year} {2021})}\BibitemShut
	{NoStop}%
	\bibitem [{\citenamefont {Matsumura}\ \emph {et~al.}(2023)\citenamefont
		{Matsumura}, \citenamefont {Fujibayashi}, \citenamefont {Kinjo},
		\citenamefont {Kitagawa}, \citenamefont {Ishida}, \citenamefont {Tokunaga},
		\citenamefont {Sakai}, \citenamefont {Kambe}, \citenamefont {Nakamura},
		\citenamefont {Shimizu}, \citenamefont {Homma}, \citenamefont {Li},
		\citenamefont {Honda},\ and\ \citenamefont {Aoki}}]{Matsumura2023}%
	\BibitemOpen
	\bibfield  {author} {\bibinfo {author} {\bibfnamefont {H.}~\bibnamefont
			{Matsumura}}, \bibinfo {author} {\bibfnamefont {H.}~\bibnamefont
			{Fujibayashi}}, \bibinfo {author} {\bibfnamefont {K.}~\bibnamefont {Kinjo}},
		\bibinfo {author} {\bibfnamefont {S.}~\bibnamefont {Kitagawa}}, \bibinfo
		{author} {\bibfnamefont {K.}~\bibnamefont {Ishida}}, \bibinfo {author}
		{\bibfnamefont {Y.}~\bibnamefont {Tokunaga}}, \bibinfo {author}
		{\bibfnamefont {H.}~\bibnamefont {Sakai}}, \bibinfo {author} {\bibfnamefont
			{S.}~\bibnamefont {Kambe}}, \bibinfo {author} {\bibfnamefont
			{A.}~\bibnamefont {Nakamura}}, \bibinfo {author} {\bibfnamefont
			{Y.}~\bibnamefont {Shimizu}}, \bibinfo {author} {\bibfnamefont
			{Y.}~\bibnamefont {Homma}}, \bibinfo {author} {\bibfnamefont
			{D.}~\bibnamefont {Li}}, \bibinfo {author} {\bibfnamefont {F.}~\bibnamefont
			{Honda}},\ and\ \bibinfo {author} {\bibfnamefont {D.}~\bibnamefont {Aoki}},\
	}\href {https://doi.org/10.7566/JPSJ.92.063701} {\bibfield  {journal}
		{\bibinfo  {journal} {J. Phys. Soc. Jpn.}\ }\textbf {\bibinfo {volume}
			{92}},\ \bibinfo {pages} {063701} (\bibinfo {year} {2023})}\BibitemShut
	{NoStop}%
	\bibitem [{\citenamefont {Ishihara}\ \emph {et~al.}(2023)\citenamefont
		{Ishihara}, \citenamefont {Kobayashi}, \citenamefont {Imamura}, \citenamefont
		{Konczykowski}, \citenamefont {Sakai}, \citenamefont {Opletal}, \citenamefont
		{Tokiwa}, \citenamefont {Haga}, \citenamefont {Hashimoto},\ and\
		\citenamefont {Shibauchi}}]{Ishihara2023}%
	\BibitemOpen
	\bibfield  {author} {\bibinfo {author} {\bibfnamefont {K.}~\bibnamefont
			{Ishihara}}, \bibinfo {author} {\bibfnamefont {M.}~\bibnamefont {Kobayashi}},
		\bibinfo {author} {\bibfnamefont {K.}~\bibnamefont {Imamura}}, \bibinfo
		{author} {\bibfnamefont {M.}~\bibnamefont {Konczykowski}}, \bibinfo {author}
		{\bibfnamefont {H.}~\bibnamefont {Sakai}}, \bibinfo {author} {\bibfnamefont
			{P.}~\bibnamefont {Opletal}}, \bibinfo {author} {\bibfnamefont
			{Y.}~\bibnamefont {Tokiwa}}, \bibinfo {author} {\bibfnamefont
			{Y.}~\bibnamefont {Haga}}, \bibinfo {author} {\bibfnamefont {K.}~\bibnamefont
			{Hashimoto}},\ and\ \bibinfo {author} {\bibfnamefont {T.}~\bibnamefont
			{Shibauchi}},\ }\href {https://doi.org/10.1103/PhysRevResearch.5.L022002}
	{\bibfield  {journal} {\bibinfo  {journal} {Phys. Rev. Res.}\ }\textbf
		{\bibinfo {volume} {5}},\ \bibinfo {pages} {L022002} (\bibinfo {year}
		{2023})}\BibitemShut {NoStop}%
	\bibitem [{\citenamefont {Iguchi}\ \emph {et~al.}(2023)\citenamefont {Iguchi},
		\citenamefont {Man}, \citenamefont {Thomas}, \citenamefont {Ronning},
		\citenamefont {Rosa},\ and\ \citenamefont {Moler}}]{Iguchi2023}%
	\BibitemOpen
	\bibfield  {author} {\bibinfo {author} {\bibfnamefont {Y.}~\bibnamefont
			{Iguchi}}, \bibinfo {author} {\bibfnamefont {H.}~\bibnamefont {Man}},
		\bibinfo {author} {\bibfnamefont {S.~M.}\ \bibnamefont {Thomas}}, \bibinfo
		{author} {\bibfnamefont {F.}~\bibnamefont {Ronning}}, \bibinfo {author}
		{\bibfnamefont {P.~F.~S.}\ \bibnamefont {Rosa}},\ and\ \bibinfo {author}
		{\bibfnamefont {K.~A.}\ \bibnamefont {Moler}},\ }\href
	{https://doi.org/10.1103/PhysRevLett.130.196003} {\bibfield  {journal}
		{\bibinfo  {journal} {Physical Review Letters}\ }\textbf {\bibinfo {volume}
			{130}},\ \bibinfo {pages} {196003} (\bibinfo {year} {2023})}\BibitemShut
	{NoStop}%
	\bibitem [{\citenamefont {Theuss}\ \emph {et~al.}(2024)\citenamefont {Theuss},
		\citenamefont {Shragai}, \citenamefont {Grissonnanche}, \citenamefont
		{Hayes}, \citenamefont {Saha}, \citenamefont {Eo}, \citenamefont {Suarez},
		\citenamefont {Shishidou}, \citenamefont {Butch}, \citenamefont {Paglione},\
		and\ \citenamefont {Ramshaw}}]{Theuss2024}%
	\BibitemOpen
	\bibfield  {author} {\bibinfo {author} {\bibfnamefont {F.}~\bibnamefont
			{Theuss}}, \bibinfo {author} {\bibfnamefont {A.}~\bibnamefont {Shragai}},
		\bibinfo {author} {\bibfnamefont {G.}~\bibnamefont {Grissonnanche}}, \bibinfo
		{author} {\bibfnamefont {I.~M.}\ \bibnamefont {Hayes}}, \bibinfo {author}
		{\bibfnamefont {S.~R.}\ \bibnamefont {Saha}}, \bibinfo {author}
		{\bibfnamefont {Y.~S.}\ \bibnamefont {Eo}}, \bibinfo {author} {\bibfnamefont
			{A.}~\bibnamefont {Suarez}}, \bibinfo {author} {\bibfnamefont
			{T.}~\bibnamefont {Shishidou}}, \bibinfo {author} {\bibfnamefont {N.~P.}\
			\bibnamefont {Butch}}, \bibinfo {author} {\bibfnamefont {J.}~\bibnamefont
			{Paglione}},\ and\ \bibinfo {author} {\bibfnamefont {B.~J.}\ \bibnamefont
			{Ramshaw}},\ }\href {https://doi.org/10.1038/s41567-024-02493-1} {\bibfield
		{journal} {\bibinfo  {journal} {Nature Physics}\ }\textbf {\bibinfo {volume}
			{20}},\ \bibinfo {pages} {1124} (\bibinfo {year} {2024})},\ \Eprint
	{https://arxiv.org/abs/2307.10938} {arXiv:2307.10938} \BibitemShut {NoStop}%
	\bibitem [{\citenamefont {Hayes}\ \emph {et~al.}(2024)\citenamefont {Hayes},
		\citenamefont {Metz}, \citenamefont {Frank}, \citenamefont {Saha},
		\citenamefont {Butch}, \citenamefont {Mishra}, \citenamefont {Hirschfeld},\
		and\ \citenamefont {Paglione}}]{Hayes2024}%
	\BibitemOpen
	\bibfield  {author} {\bibinfo {author} {\bibfnamefont {I.~M.}\ \bibnamefont
			{Hayes}}, \bibinfo {author} {\bibfnamefont {T.~E.}\ \bibnamefont {Metz}},
		\bibinfo {author} {\bibfnamefont {C.~E.}\ \bibnamefont {Frank}}, \bibinfo
		{author} {\bibfnamefont {S.~R.}\ \bibnamefont {Saha}}, \bibinfo {author}
		{\bibfnamefont {N.~P.}\ \bibnamefont {Butch}}, \bibinfo {author}
		{\bibfnamefont {V.}~\bibnamefont {Mishra}}, \bibinfo {author} {\bibfnamefont
			{P.~J.}\ \bibnamefont {Hirschfeld}},\ and\ \bibinfo {author} {\bibfnamefont
			{J.}~\bibnamefont {Paglione}},\ }\href {https://arxiv.org/abs/2402.19353}
	{\bibinfo {title} {Robust nodal behavior in the thermal conductivity of
			superconducting ute$_2$}} (\bibinfo {year} {2024}),\ \Eprint
	{https://arxiv.org/abs/2402.19353} {arXiv:2402.19353 [cond-mat.supr-con]}
	\BibitemShut {NoStop}%
	\bibitem [{Sup()}]{SuppMat}%
	\BibitemOpen
	\href@noop {} {}\bibinfo {note} {Supplemental Material contains information
		on the ac specific heat technique under high pressure; it presents additional
		data on samples A, B, and C; gives details on the fitting procedure used to
		determine the superconducting transition temperatures, the width of the
		superconducting transitions and the height of the specific heat jump at the
		transition. }
		\BibitemShut {Stop}%
	\bibitem [{\citenamefont {Knebel}\ \emph {et~al.}(2020)\citenamefont {Knebel},
		\citenamefont {Kimata}, \citenamefont {Vali{\v{s}}ka}, \citenamefont {Honda},
		\citenamefont {Li}, \citenamefont {Braithwaite}, \citenamefont {Lapertot},
		\citenamefont {Knafo}, \citenamefont {Pourret}, \citenamefont {Sato},
		\citenamefont {Shimizu}, \citenamefont {Kihara}, \citenamefont {Brison},
		\citenamefont {Flouquet},\ and\ \citenamefont {Aoki}}]{Knebel2020}%
	\BibitemOpen
	\bibfield  {author} {\bibinfo {author} {\bibfnamefont {G.}~\bibnamefont
			{Knebel}}, \bibinfo {author} {\bibfnamefont {M.}~\bibnamefont {Kimata}},
		\bibinfo {author} {\bibfnamefont {M.}~\bibnamefont {Vali{\v{s}}ka}}, \bibinfo
		{author} {\bibfnamefont {F.}~\bibnamefont {Honda}}, \bibinfo {author}
		{\bibfnamefont {D.~X.}\ \bibnamefont {Li}}, \bibinfo {author} {\bibfnamefont
			{D.}~\bibnamefont {Braithwaite}}, \bibinfo {author} {\bibfnamefont
			{G.}~\bibnamefont {Lapertot}}, \bibinfo {author} {\bibfnamefont
			{W.}~\bibnamefont {Knafo}}, \bibinfo {author} {\bibfnamefont
			{A.}~\bibnamefont {Pourret}}, \bibinfo {author} {\bibfnamefont {Y.~J.}\
			\bibnamefont {Sato}}, \bibinfo {author} {\bibfnamefont {Y.}~\bibnamefont
			{Shimizu}}, \bibinfo {author} {\bibfnamefont {T.}~\bibnamefont {Kihara}},
		\bibinfo {author} {\bibfnamefont {J.~P.}\ \bibnamefont {Brison}}, \bibinfo
		{author} {\bibfnamefont {J.}~\bibnamefont {Flouquet}},\ and\ \bibinfo
		{author} {\bibfnamefont {D.}~\bibnamefont {Aoki}},\ }\href
	{https://doi.org/10.7566/JPSJ.89.053707} {\bibfield  {journal} {\bibinfo
			{journal} {J. Phys. Soc. Jpn.}\ }\textbf {\bibinfo {volume} {89}},\ \bibinfo
		{pages} {1} (\bibinfo {year} {2020})}\BibitemShut {NoStop}%
	\bibitem [{\citenamefont {Knebel}\ \emph {et~al.}(2024)\citenamefont {Knebel},
		\citenamefont {Pourret}, \citenamefont {Rousseau}, \citenamefont {Marquardt},
		\citenamefont {Braithwaite}, \citenamefont {Honda}, \citenamefont {Aoki},
		\citenamefont {Lapertot}, \citenamefont {Knafo}, \citenamefont {Seyfarth},
		\citenamefont {Brison},\ and\ \citenamefont {Flouquet}}]{Knebel2024}%
	\BibitemOpen
	\bibfield  {author} {\bibinfo {author} {\bibfnamefont {G.}~\bibnamefont
			{Knebel}}, \bibinfo {author} {\bibfnamefont {A.}~\bibnamefont {Pourret}},
		\bibinfo {author} {\bibfnamefont {S.}~\bibnamefont {Rousseau}}, \bibinfo
		{author} {\bibfnamefont {N.}~\bibnamefont {Marquardt}}, \bibinfo {author}
		{\bibfnamefont {D.}~\bibnamefont {Braithwaite}}, \bibinfo {author}
		{\bibfnamefont {F.}~\bibnamefont {Honda}}, \bibinfo {author} {\bibfnamefont
			{D.}~\bibnamefont {Aoki}}, \bibinfo {author} {\bibfnamefont {G.}~\bibnamefont
			{Lapertot}}, \bibinfo {author} {\bibfnamefont {W.}~\bibnamefont {Knafo}},
		\bibinfo {author} {\bibfnamefont {G.}~\bibnamefont {Seyfarth}}, \bibinfo
		{author} {\bibfnamefont {J.-P.}\ \bibnamefont {Brison}},\ and\ \bibinfo
		{author} {\bibfnamefont {J.}~\bibnamefont {Flouquet}},\ }\href
	{https://doi.org/10.1103/physrevb.109.155103} {\bibfield  {journal} {\bibinfo
			{journal} {Phys. Rev. B}\ }\textbf {\bibinfo {volume} {109}},\ \bibinfo
		{pages} {155103} (\bibinfo {year} {2024})}\BibitemShut {NoStop}%
	\bibitem [{\citenamefont {Imajo}\ \emph {et~al.}(2019)\citenamefont {Imajo},
		\citenamefont {Kohama}, \citenamefont {Miyake}, \citenamefont {Dong},
		\citenamefont {Tokunaga}, \citenamefont {Flouquet}, \citenamefont {Kindo},\
		and\ \citenamefont {Aoki}}]{Imajo2019}%
	\BibitemOpen
	\bibfield  {author} {\bibinfo {author} {\bibfnamefont {S.}~\bibnamefont
			{Imajo}}, \bibinfo {author} {\bibfnamefont {Y.}~\bibnamefont {Kohama}},
		\bibinfo {author} {\bibfnamefont {A.}~\bibnamefont {Miyake}}, \bibinfo
		{author} {\bibfnamefont {C.}~\bibnamefont {Dong}}, \bibinfo {author}
		{\bibfnamefont {M.}~\bibnamefont {Tokunaga}}, \bibinfo {author}
		{\bibfnamefont {J.}~\bibnamefont {Flouquet}}, \bibinfo {author}
		{\bibfnamefont {K.}~\bibnamefont {Kindo}},\ and\ \bibinfo {author}
		{\bibfnamefont {D.}~\bibnamefont {Aoki}},\ }\href
	{https://doi.org/10.7566/JPSJ.88.083705} {\bibfield  {journal} {\bibinfo
			{journal} {J. Phys. Soc. Jpn.}\ }\textbf {\bibinfo {volume} {88}},\ \bibinfo
		{pages} {083705} (\bibinfo {year} {2019})}\BibitemShut {NoStop}%
	\bibitem [{\citenamefont {Miyake}\ \emph {et~al.}(2021)\citenamefont {Miyake},
		\citenamefont {Shimizu}, \citenamefont {Sato}, \citenamefont {Li},
		\citenamefont {Nakamura}, \citenamefont {Homma}, \citenamefont {Honda},
		\citenamefont {Flouquet}, \citenamefont {Tokunaga},\ and\ \citenamefont
		{Aoki}}]{Miyake2021b}%
	\BibitemOpen
	\bibfield  {author} {\bibinfo {author} {\bibfnamefont {A.}~\bibnamefont
			{Miyake}}, \bibinfo {author} {\bibfnamefont {Y.}~\bibnamefont {Shimizu}},
		\bibinfo {author} {\bibfnamefont {Y.~J.}\ \bibnamefont {Sato}}, \bibinfo
		{author} {\bibfnamefont {D.}~\bibnamefont {Li}}, \bibinfo {author}
		{\bibfnamefont {A.}~\bibnamefont {Nakamura}}, \bibinfo {author}
		{\bibfnamefont {Y.}~\bibnamefont {Homma}}, \bibinfo {author} {\bibfnamefont
			{F.}~\bibnamefont {Honda}}, \bibinfo {author} {\bibfnamefont
			{J.}~\bibnamefont {Flouquet}}, \bibinfo {author} {\bibfnamefont
			{M.}~\bibnamefont {Tokunaga}},\ and\ \bibinfo {author} {\bibfnamefont
			{D.}~\bibnamefont {Aoki}},\ }\href {https://doi.org/10.7566/JPSJ.90.103702}
	{\bibfield  {journal} {\bibinfo  {journal} {J. Phys. Soc. Jpn.}\ }\textbf
		{\bibinfo {volume} {90}},\ \bibinfo {pages} {103702} (\bibinfo {year}
		{2021})}\BibitemShut {NoStop}%
	\bibitem [{\citenamefont {Knafo}\ \emph {et~al.}(2020)\citenamefont {Knafo},
		\citenamefont {Nardone}, \citenamefont {Valiska}, \citenamefont {Zitouni},
		\citenamefont {Lapertot}, \citenamefont {Aoki}, \citenamefont {Knebel},\ and\
		\citenamefont {Braithwaite}}]{Knafo2020}%
	\BibitemOpen
	\bibfield  {author} {\bibinfo {author} {\bibfnamefont {W.}~\bibnamefont
			{Knafo}}, \bibinfo {author} {\bibfnamefont {M.}~\bibnamefont {Nardone}},
		\bibinfo {author} {\bibfnamefont {M.}~\bibnamefont {Valiska}}, \bibinfo
		{author} {\bibfnamefont {A.}~\bibnamefont {Zitouni}}, \bibinfo {author}
		{\bibfnamefont {G.}~\bibnamefont {Lapertot}}, \bibinfo {author}
		{\bibfnamefont {D.}~\bibnamefont {Aoki}}, \bibinfo {author} {\bibfnamefont
			{G.}~\bibnamefont {Knebel}},\ and\ \bibinfo {author} {\bibfnamefont
			{D.}~\bibnamefont {Braithwaite}},\ }\href
	{https://doi.org/10.1038/s42005-021-00545-z} {\bibfield  {journal} {\bibinfo
			{journal} {Commun Phys}\ }\textbf {\bibinfo {volume} {4}},\ \bibinfo {pages}
		{40} (\bibinfo {year} {2020})}\BibitemShut {NoStop}%
	\bibitem [{\citenamefont {Yip}\ \emph {et~al.}(1991)\citenamefont {Yip},
		\citenamefont {Li},\ and\ \citenamefont {Kumar}}]{Yip1991}%
	\BibitemOpen
	\bibfield  {author} {\bibinfo {author} {\bibfnamefont {S.~K.}\ \bibnamefont
			{Yip}}, \bibinfo {author} {\bibfnamefont {T.}~\bibnamefont {Li}},\ and\
		\bibinfo {author} {\bibfnamefont {P.}~\bibnamefont {Kumar}},\ }\href
	{https://doi.org/10.1103/PhysRevB.43.2742} {\bibfield  {journal} {\bibinfo
			{journal} {Phys. Rev. B}\ }\textbf {\bibinfo {volume} {43}},\ \bibinfo
		{pages} {2742} (\bibinfo {year} {1991})}\BibitemShut {NoStop}%
	\bibitem [{\citenamefont {Kinjo}\ \emph
		{et~al.}(2023{\natexlab{b}})\citenamefont {Kinjo}, \citenamefont
		{Fujibayashi}, \citenamefont {Matsumura}, \citenamefont {Hori}, \citenamefont
		{Kitagawa}, \citenamefont {Ishida}, \citenamefont {Tokunaga}, \citenamefont
		{Sakai}, \citenamefont {Kambe}, \citenamefont {Nakamura}, \citenamefont
		{Shimizu}, \citenamefont {Homma}, \citenamefont {Li}, \citenamefont {Honda},\
		and\ \citenamefont {Aoki}}]{Kinjo2023}%
	\BibitemOpen
	\bibfield  {author} {\bibinfo {author} {\bibfnamefont {K.}~\bibnamefont
			{Kinjo}}, \bibinfo {author} {\bibfnamefont {H.}~\bibnamefont {Fujibayashi}},
		\bibinfo {author} {\bibfnamefont {H.}~\bibnamefont {Matsumura}}, \bibinfo
		{author} {\bibfnamefont {F.}~\bibnamefont {Hori}}, \bibinfo {author}
		{\bibfnamefont {S.}~\bibnamefont {Kitagawa}}, \bibinfo {author}
		{\bibfnamefont {K.}~\bibnamefont {Ishida}}, \bibinfo {author} {\bibfnamefont
			{Y.}~\bibnamefont {Tokunaga}}, \bibinfo {author} {\bibfnamefont
			{H.}~\bibnamefont {Sakai}}, \bibinfo {author} {\bibfnamefont
			{S.}~\bibnamefont {Kambe}}, \bibinfo {author} {\bibfnamefont
			{A.}~\bibnamefont {Nakamura}}, \bibinfo {author} {\bibfnamefont
			{Y.}~\bibnamefont {Shimizu}}, \bibinfo {author} {\bibfnamefont
			{Y.}~\bibnamefont {Homma}}, \bibinfo {author} {\bibfnamefont
			{D.}~\bibnamefont {Li}}, \bibinfo {author} {\bibfnamefont {F.}~\bibnamefont
			{Honda}},\ and\ \bibinfo {author} {\bibfnamefont {D.}~\bibnamefont {Aoki}},\
	}\href {https://doi.org/10.1126/sciadv.adg2736} {\bibfield  {journal}
		{\bibinfo  {journal} {Sci. Adv.}\ }\textbf {\bibinfo {volume} {9}},\ \bibinfo
		{pages} {eadg2736} (\bibinfo {year} {2023}{\natexlab{b}})}\BibitemShut
	{NoStop}%
	\bibitem [{\citenamefont {Xu}\ \emph {et~al.}(2019)\citenamefont {Xu},
		\citenamefont {Sheng},\ and\ \citenamefont {Yang}}]{Xu2019}%
	\BibitemOpen
	\bibfield  {author} {\bibinfo {author} {\bibfnamefont {Y.}~\bibnamefont
			{Xu}}, \bibinfo {author} {\bibfnamefont {Y.}~\bibnamefont {Sheng}},\ and\
		\bibinfo {author} {\bibfnamefont {Y.-f.}\ \bibnamefont {Yang}},\ }\href
	{https://doi.org/10.1103/PhysRevLett.123.217002} {\bibfield  {journal}
		{\bibinfo  {journal} {Phys. Rev. Lett.}\ }\textbf {\bibinfo {volume} {123}},\
		\bibinfo {pages} {217002} (\bibinfo {year} {2019})},\ \Eprint
	{https://arxiv.org/abs/1908.07396} {arXiv:1908.07396} \BibitemShut {NoStop}%
	\bibitem [{\citenamefont {Shishidou}\ \emph {et~al.}(2021)\citenamefont
		{Shishidou}, \citenamefont {Suh}, \citenamefont {Brydon}, \citenamefont
		{Weinert},\ and\ \citenamefont {Agterberg}}]{Shishidou2021}%
	\BibitemOpen
	\bibfield  {author} {\bibinfo {author} {\bibfnamefont {T.}~\bibnamefont
			{Shishidou}}, \bibinfo {author} {\bibfnamefont {H.~G.}\ \bibnamefont {Suh}},
		\bibinfo {author} {\bibfnamefont {P.~M.~R.}\ \bibnamefont {Brydon}}, \bibinfo
		{author} {\bibfnamefont {M.}~\bibnamefont {Weinert}},\ and\ \bibinfo {author}
		{\bibfnamefont {D.~F.}\ \bibnamefont {Agterberg}},\ }\href
	{https://doi.org/10.1103/PhysRevB.103.104504} {\bibfield  {journal} {\bibinfo
			{journal} {Phys. Rev. B}\ }\textbf {\bibinfo {volume} {103}},\ \bibinfo
		{pages} {104504} (\bibinfo {year} {2021})}\BibitemShut {NoStop}%
	\bibitem [{\citenamefont {Ishizuka}\ and\ \citenamefont
		{Yanase}(2021)}]{Ishizuka2021}%
	\BibitemOpen
	\bibfield  {author} {\bibinfo {author} {\bibfnamefont {J.}~\bibnamefont
			{Ishizuka}}\ and\ \bibinfo {author} {\bibfnamefont {Y.}~\bibnamefont
			{Yanase}},\ }\href {https://doi.org/10.1103/PhysRevB.103.094504} {\bibfield
		{journal} {\bibinfo  {journal} {Phys. Rev. B}\ }\textbf {\bibinfo {volume}
			{103}},\ \bibinfo {pages} {094504} (\bibinfo {year} {2021})}\BibitemShut
	{NoStop}%
	\bibitem [{\citenamefont {Kreisel}\ \emph {et~al.}(2022)\citenamefont
		{Kreisel}, \citenamefont {Quan},\ and\ \citenamefont
		{Hirschfeld}}]{KreiselPRB2022}%
	\BibitemOpen
	\bibfield  {author} {\bibinfo {author} {\bibfnamefont {A.}~\bibnamefont
			{Kreisel}}, \bibinfo {author} {\bibfnamefont {Y.}~\bibnamefont {Quan}},\ and\
		\bibinfo {author} {\bibfnamefont {P.~J.}\ \bibnamefont {Hirschfeld}},\ }\href
	{https://doi.org/10.1103/physrevb.105.104507} {\bibfield  {journal} {\bibinfo
			{journal} {Physical Review B}\ }\textbf {\bibinfo {volume} {105}},\ \bibinfo
		{pages} {104507} (\bibinfo {year} {2022})}\BibitemShut {NoStop}%
	\bibitem [{\citenamefont {Tei}\ \emph {et~al.}(2024)\citenamefont {Tei},
		\citenamefont {Mizushima},\ and\ \citenamefont {Fujimoto}}]{TeiPRB2024}%
	\BibitemOpen
	\bibfield  {author} {\bibinfo {author} {\bibfnamefont {J.}~\bibnamefont
			{Tei}}, \bibinfo {author} {\bibfnamefont {T.}~\bibnamefont {Mizushima}},\
		and\ \bibinfo {author} {\bibfnamefont {S.}~\bibnamefont {Fujimoto}},\ }\href
	{https://doi.org/10.1103/PhysRevB.109.064516} {\bibfield  {journal} {\bibinfo
			{journal} {Phys. Rev. B}\ }\textbf {\bibinfo {volume} {109}},\ \bibinfo
		{pages} {064516} (\bibinfo {year} {2024})}\BibitemShut {NoStop}%
	\bibitem [{\citenamefont {Wilhelm}\ \emph {et~al.}(2023)\citenamefont
		{Wilhelm}, \citenamefont {Sanchez}, \citenamefont {Braithwaite},
		\citenamefont {Knebel}, \citenamefont {Lapertot},\ and\ \citenamefont
		{Rogalev}}]{Wilhelm2023}%
	\BibitemOpen
	\bibfield  {author} {\bibinfo {author} {\bibfnamefont {F.}~\bibnamefont
			{Wilhelm}}, \bibinfo {author} {\bibfnamefont {J.-P.}\ \bibnamefont
			{Sanchez}}, \bibinfo {author} {\bibfnamefont {D.}~\bibnamefont
			{Braithwaite}}, \bibinfo {author} {\bibfnamefont {G.}~\bibnamefont {Knebel}},
		\bibinfo {author} {\bibfnamefont {G.}~\bibnamefont {Lapertot}},\ and\
		\bibinfo {author} {\bibfnamefont {A.}~\bibnamefont {Rogalev}},\ }\href
	{https://doi.org/10.1038/s42005-023-01220-1} {\bibfield  {journal} {\bibinfo
			{journal} {Commun Phys}\ }\textbf {\bibinfo {volume} {6}},\ \bibinfo {pages}
		{96} (\bibinfo {year} {2023})}\BibitemShut {NoStop}%
	\bibitem [{\citenamefont {Li}\ \emph {et~al.}(2021)\citenamefont {Li},
		\citenamefont {Nakamura}, \citenamefont {Honda}, \citenamefont {Sato},
		\citenamefont {Homma}, \citenamefont {Shimizu}, \citenamefont {Ishizuka},
		\citenamefont {Yanase}, \citenamefont {Knebel}, \citenamefont {Flouquet},\
		and\ \citenamefont {Aoki}}]{Li2021}%
	\BibitemOpen
	\bibfield  {author} {\bibinfo {author} {\bibfnamefont {D.}~\bibnamefont
			{Li}}, \bibinfo {author} {\bibfnamefont {A.}~\bibnamefont {Nakamura}},
		\bibinfo {author} {\bibfnamefont {F.}~\bibnamefont {Honda}}, \bibinfo
		{author} {\bibfnamefont {Y.~J.}\ \bibnamefont {Sato}}, \bibinfo {author}
		{\bibfnamefont {Y.}~\bibnamefont {Homma}}, \bibinfo {author} {\bibfnamefont
			{Y.}~\bibnamefont {Shimizu}}, \bibinfo {author} {\bibfnamefont
			{J.}~\bibnamefont {Ishizuka}}, \bibinfo {author} {\bibfnamefont
			{Y.}~\bibnamefont {Yanase}}, \bibinfo {author} {\bibfnamefont
			{G.}~\bibnamefont {Knebel}}, \bibinfo {author} {\bibfnamefont
			{J.}~\bibnamefont {Flouquet}},\ and\ \bibinfo {author} {\bibfnamefont
			{D.}~\bibnamefont {Aoki}},\ }\href {https://doi.org/10.7566/JPSJ.90.073703}
	{\bibfield  {journal} {\bibinfo  {journal} {J. Phys. Soc. Jpn.}\ }\textbf
		{\bibinfo {volume} {90}},\ \bibinfo {pages} {073703} (\bibinfo {year}
		{2021})}\BibitemShut {NoStop}%
	\bibitem [{\citenamefont {Hakuno}\ \emph {et~al.}(2024)\citenamefont {Hakuno},
		\citenamefont {Nogaki},\ and\ \citenamefont {Yanase}}]{HakunoPRB2024}%
	\BibitemOpen
	\bibfield  {author} {\bibinfo {author} {\bibfnamefont {R.}~\bibnamefont
			{Hakuno}}, \bibinfo {author} {\bibfnamefont {K.}~\bibnamefont {Nogaki}},\
		and\ \bibinfo {author} {\bibfnamefont {Y.}~\bibnamefont {Yanase}},\ }\href
	{https://doi.org/10.1103/PhysRevB.109.104509} {\bibfield  {journal} {\bibinfo
			{journal} {Phys. Rev. B}\ }\textbf {\bibinfo {volume} {109}},\ \bibinfo
		{pages} {104509} (\bibinfo {year} {2024})}\BibitemShut {NoStop}%
	\bibitem [{\citenamefont {Knafo}\ \emph {et~al.}(2023)\citenamefont {Knafo},
		\citenamefont {Thebault}, \citenamefont {Manuel}, \citenamefont {Khalyavin},
		\citenamefont {Orlandi}, \citenamefont {Ressouche}, \citenamefont {Beauvois},
		\citenamefont {Lapertot}, \citenamefont {Kaneko}, \citenamefont {Aoki},
		\citenamefont {Braithwaite}, \citenamefont {Knebel},\ and\ \citenamefont
		{Raymond}}]{Knafo2023}%
	\BibitemOpen
	\bibfield  {author} {\bibinfo {author} {\bibfnamefont {W.}~\bibnamefont
			{Knafo}}, \bibinfo {author} {\bibfnamefont {T.}~\bibnamefont {Thebault}},
		\bibinfo {author} {\bibfnamefont {P.}~\bibnamefont {Manuel}}, \bibinfo
		{author} {\bibfnamefont {D.~D.}\ \bibnamefont {Khalyavin}}, \bibinfo {author}
		{\bibfnamefont {F.}~\bibnamefont {Orlandi}}, \bibinfo {author} {\bibfnamefont
			{E.}~\bibnamefont {Ressouche}}, \bibinfo {author} {\bibfnamefont
			{K.}~\bibnamefont {Beauvois}}, \bibinfo {author} {\bibfnamefont
			{G.}~\bibnamefont {Lapertot}}, \bibinfo {author} {\bibfnamefont
			{K.}~\bibnamefont {Kaneko}}, \bibinfo {author} {\bibfnamefont
			{D.}~\bibnamefont {Aoki}}, \bibinfo {author} {\bibfnamefont {D.}~\bibnamefont
			{Braithwaite}}, \bibinfo {author} {\bibfnamefont {G.}~\bibnamefont
			{Knebel}},\ and\ \bibinfo {author} {\bibfnamefont {S.}~\bibnamefont
			{Raymond}},\ }\href@noop {} {\bibinfo {title} {{Incommensurate
				antiferromagnetism in UTe$_2$ under pressure}}} (\bibinfo {year} {2023}),\
	\Eprint {https://arxiv.org/abs/2311.05455} {arXiv:2311.05455
		[cond-mat.str-el]} \BibitemShut {NoStop}%
	\bibitem [{\citenamefont {Duan}\ \emph {et~al.}(2020)\citenamefont {Duan},
		\citenamefont {Sasmal}, \citenamefont {Maple}, \citenamefont {Podlesnyak},
		\citenamefont {Zhu}, \citenamefont {Si},\ and\ \citenamefont
		{Dai}}]{Duan2020}%
	\BibitemOpen
	\bibfield  {author} {\bibinfo {author} {\bibfnamefont {C.}~\bibnamefont
			{Duan}}, \bibinfo {author} {\bibfnamefont {K.}~\bibnamefont {Sasmal}},
		\bibinfo {author} {\bibfnamefont {M.~B.}\ \bibnamefont {Maple}}, \bibinfo
		{author} {\bibfnamefont {A.}~\bibnamefont {Podlesnyak}}, \bibinfo {author}
		{\bibfnamefont {J.-X.}\ \bibnamefont {Zhu}}, \bibinfo {author} {\bibfnamefont
			{Q.}~\bibnamefont {Si}},\ and\ \bibinfo {author} {\bibfnamefont
			{P.}~\bibnamefont {Dai}},\ }\href
	{https://doi.org/10.1103/PhysRevLett.125.237003} {\bibfield  {journal}
		{\bibinfo  {journal} {Phys. Rev. Lett.}\ }\textbf {\bibinfo {volume} {125}},\
		\bibinfo {pages} {237003} (\bibinfo {year} {2020})}\BibitemShut {NoStop}%
	\bibitem [{\citenamefont {Knafo}\ \emph {et~al.}(2021)\citenamefont {Knafo},
		\citenamefont {Knebel}, \citenamefont {Steffens}, \citenamefont {Kaneko},
		\citenamefont {Rosuel}, \citenamefont {Brison}, \citenamefont {Flouquet},
		\citenamefont {Aoki}, \citenamefont {Lapertot},\ and\ \citenamefont
		{Raymond}}]{Knafo2021}%
	\BibitemOpen
	\bibfield  {author} {\bibinfo {author} {\bibfnamefont {W.}~\bibnamefont
			{Knafo}}, \bibinfo {author} {\bibfnamefont {G.}~\bibnamefont {Knebel}},
		\bibinfo {author} {\bibfnamefont {P.}~\bibnamefont {Steffens}}, \bibinfo
		{author} {\bibfnamefont {K.}~\bibnamefont {Kaneko}}, \bibinfo {author}
		{\bibfnamefont {A.}~\bibnamefont {Rosuel}}, \bibinfo {author} {\bibfnamefont
			{J.-P.}\ \bibnamefont {Brison}}, \bibinfo {author} {\bibfnamefont
			{J.}~\bibnamefont {Flouquet}}, \bibinfo {author} {\bibfnamefont
			{D.}~\bibnamefont {Aoki}}, \bibinfo {author} {\bibfnamefont {G.}~\bibnamefont
			{Lapertot}},\ and\ \bibinfo {author} {\bibfnamefont {S.}~\bibnamefont
			{Raymond}},\ }\href {https://doi.org/10.1103/PhysRevB.104.L100409} {\bibfield
		{journal} {\bibinfo  {journal} {Phys. Rev. B}\ }\textbf {\bibinfo {volume}
			{104}},\ \bibinfo {pages} {L100409} (\bibinfo {year} {2021})}\BibitemShut
	{NoStop}%

\end{thebibliography}

\begin{thebibliography}{6}%
	\makeatletter
	\providecommand \@ifxundefined [1]{%
		\@ifx{#1\undefined}
	}%
	\providecommand \@ifnum [1]{%
		\ifnum #1\expandafter \@firstoftwo
		\else \expandafter \@secondoftwo
		\fi
	}%
	\providecommand \@ifx [1]{%
		\ifx #1\expandafter \@firstoftwo
		\else \expandafter \@secondoftwo
		\fi
	}%
	\providecommand \natexlab [1]{#1}%
	\providecommand \enquote  [1]{``#1''}%
	\providecommand \bibnamefont  [1]{#1}%
	\providecommand \bibfnamefont [1]{#1}%
	\providecommand \citenamefont [1]{#1}%
	\providecommand \href@noop [0]{\@secondoftwo}%
	\providecommand \href [0]{\begingroup \@sanitize@url \@href}%
	\providecommand \@href[1]{\@@startlink{#1}\@@href}%
	\providecommand \@@href[1]{\endgroup#1\@@endlink}%
	\providecommand \@sanitize@url [0]{\catcode `\\12\catcode `\$12\catcode
		`\&12\catcode `\#12\catcode `\^12\catcode `\_12\catcode `\%12\relax}%
	\providecommand \@@startlink[1]{}%
	\providecommand \@@endlink[0]{}%
	\providecommand \url  [0]{\begingroup\@sanitize@url \@url }%
	\providecommand \@url [1]{\endgroup\@href {#1}{\urlprefix }}%
	\providecommand \urlprefix  [0]{URL }%
	\providecommand \Eprint [0]{\href }%
	\providecommand \doibase [0]{https://doi.org/}%
	\providecommand \selectlanguage [0]{\@gobble}%
	\providecommand \bibinfo  [0]{\@secondoftwo}%
	\providecommand \bibfield  [0]{\@secondoftwo}%
	\providecommand \translation [1]{[#1]}%
	\providecommand \BibitemOpen [0]{}%
	\providecommand \bibitemStop [0]{}%
	\providecommand \bibitemNoStop [0]{.\EOS\space}%
	\providecommand \EOS [0]{\spacefactor3000\relax}%
	\providecommand \BibitemShut  [1]{\csname bibitem#1\endcsname}%
	\let\auto@bib@innerbib\@empty
	\bibitem [{\citenamefont {Sakai}\ \emph {et~al.}(2022)\citenamefont {Sakai},
		\citenamefont {Opletal}, \citenamefont {Tokiwa}, \citenamefont {Yamamoto},
		\citenamefont {Tokunaga}, \citenamefont {Kambe},\ and\ \citenamefont
		{Haga}}]{Sakai2022}%
	\BibitemOpen
	\bibfield  {author} {\bibinfo {author} {\bibfnamefont {H.}~\bibnamefont
			{Sakai}}, \bibinfo {author} {\bibfnamefont {P.}~\bibnamefont {Opletal}},
		\bibinfo {author} {\bibfnamefont {Y.}~\bibnamefont {Tokiwa}}, \bibinfo
		{author} {\bibfnamefont {E.}~\bibnamefont {Yamamoto}}, \bibinfo {author}
		{\bibfnamefont {Y.}~\bibnamefont {Tokunaga}}, \bibinfo {author}
		{\bibfnamefont {S.}~\bibnamefont {Kambe}},\ and\ \bibinfo {author}
		{\bibfnamefont {Y.}~\bibnamefont {Haga}},\ }\bibfield  {title} {\bibinfo
		{title} {{Single crystal growth of superconducting ${\mathrm{UTe}}_{2}$ by
				molten salt flux method}},\ }\href
	{https://doi.org/10.1103/PhysRevMaterials.6.073401} {\bibfield  {journal}
		{\bibinfo  {journal} {Phys. Rev. Mater.}\ }\textbf {\bibinfo {volume} {6}},\
		\bibinfo {pages} {073401} (\bibinfo {year} {2022})}\BibitemShut {NoStop}%
	\bibitem [{\citenamefont {Aoki}(2024)}]{Aoki2024}%
	\BibitemOpen
	\bibfield  {author} {\bibinfo {author} {\bibfnamefont {D.}~\bibnamefont
			{Aoki}},\ }\bibfield  {title} {\bibinfo {title} {Molten salt flux liquid
			transport method for ultra clean single crystals {UTe$_2$}},\ }\href
	{https://doi.org/10.7566/JPSJ.93.043703} {\bibfield  {journal} {\bibinfo
			{journal} {J.~Phys.~Soc.~Jpn.}\ }\textbf {\bibinfo {volume} {93}},\ \bibinfo
		{pages} {043703} (\bibinfo {year} {2024})}\BibitemShut {NoStop}%
	\bibitem [{\citenamefont {Xiang}\ \emph {et~al.}(2020)\citenamefont {Xiang},
		\citenamefont {Gati}, \citenamefont {Bud’ko}, \citenamefont {Ribeiro},
		\citenamefont {Ata}, \citenamefont {Tutsch}, \citenamefont {Lang},\ and\
		\citenamefont {Canfield}}]{xiang_characterization_2020}%
	\BibitemOpen
	\bibfield  {author} {\bibinfo {author} {\bibfnamefont {L.}~\bibnamefont
			{Xiang}}, \bibinfo {author} {\bibfnamefont {E.}~\bibnamefont {Gati}},
		\bibinfo {author} {\bibfnamefont {S.~L.}\ \bibnamefont {Bud’ko}}, \bibinfo
		{author} {\bibfnamefont {R.~A.}\ \bibnamefont {Ribeiro}}, \bibinfo {author}
		{\bibfnamefont {A.}~\bibnamefont {Ata}}, \bibinfo {author} {\bibfnamefont
			{U.}~\bibnamefont {Tutsch}}, \bibinfo {author} {\bibfnamefont
			{M.}~\bibnamefont {Lang}},\ and\ \bibinfo {author} {\bibfnamefont {P.~C.}\
			\bibnamefont {Canfield}},\ }\bibfield  {title} {{\bibinfo
			{title} {Characterization of the pressure coefficient of manganin and
				temperature evolution of pressure in piston-cylinder cells}},\ }\href
	{https://doi.org/10.1063/5.0022650} {\bibfield  {journal} {\bibinfo
			{journal} {Review of Scientific Instruments}\ }\textbf {\bibinfo {volume}
			{91}},\ \bibinfo {pages} {095103} (\bibinfo {year} {2020})}\BibitemShut
	{NoStop}%
	\bibitem [{\citenamefont {Gati}\ \emph {et~al.}(2019)\citenamefont {Gati},
		\citenamefont {Drachuck}, \citenamefont {Xiang}, \citenamefont {Wang},
		\citenamefont {Bud’ko},\ and\ \citenamefont {Canfield}}]{gati_use_2019}%
	\BibitemOpen
	\bibfield  {author} {\bibinfo {author} {\bibfnamefont {E.}~\bibnamefont
			{Gati}}, \bibinfo {author} {\bibfnamefont {G.}~\bibnamefont {Drachuck}},
		\bibinfo {author} {\bibfnamefont {L.}~\bibnamefont {Xiang}}, \bibinfo
		{author} {\bibfnamefont {L.-L.}\ \bibnamefont {Wang}}, \bibinfo {author}
		{\bibfnamefont {S.~L.}\ \bibnamefont {Bud’ko}},\ and\ \bibinfo {author}
		{\bibfnamefont {P.~C.}\ \bibnamefont {Canfield}},\ }\bibfield  {title}
	{{\bibinfo {title} {Use of {Cernox} thermometers in {AC}
				specific heat measurements under pressure}},\ }\href
	{https://doi.org/10.1063/1.5084730} {\bibfield  {journal} {\bibinfo
			{journal} {Review of Scientific Instruments}\ }\textbf {\bibinfo {volume}
			{90}},\ \bibinfo {pages} {023911} (\bibinfo {year} {2019})}\BibitemShut
	{NoStop}%
	\bibitem [{\citenamefont {Rosuel}\ \emph {et~al.}(2023)\citenamefont {Rosuel},
		\citenamefont {Marcenat}, \citenamefont {Knebel}, \citenamefont {Klein},
		\citenamefont {Pourret}, \citenamefont {Marquardt}, \citenamefont {Niu},
		\citenamefont {Rousseau}, \citenamefont {Demuer}, \citenamefont {Seyfarth},
		\citenamefont {Lapertot}, \citenamefont {Aoki}, \citenamefont {Braithwaite},
		\citenamefont {Flouquet},\ and\ \citenamefont {Brison}}]{Rosuel2023}%
	\BibitemOpen
	\bibfield  {author} {\bibinfo {author} {\bibfnamefont {A.}~\bibnamefont
			{Rosuel}}, \bibinfo {author} {\bibfnamefont {C.}~\bibnamefont {Marcenat}},
		\bibinfo {author} {\bibfnamefont {G.}~\bibnamefont {Knebel}}, \bibinfo
		{author} {\bibfnamefont {T.}~\bibnamefont {Klein}}, \bibinfo {author}
		{\bibfnamefont {A.}~\bibnamefont {Pourret}}, \bibinfo {author} {\bibfnamefont
			{N.}~\bibnamefont {Marquardt}}, \bibinfo {author} {\bibfnamefont
			{Q.}~\bibnamefont {Niu}}, \bibinfo {author} {\bibfnamefont {S.}~\bibnamefont
			{Rousseau}}, \bibinfo {author} {\bibfnamefont {A.}~\bibnamefont {Demuer}},
		\bibinfo {author} {\bibfnamefont {G.}~\bibnamefont {Seyfarth}}, \bibinfo
		{author} {\bibfnamefont {G.}~\bibnamefont {Lapertot}}, \bibinfo {author}
		{\bibfnamefont {D.}~\bibnamefont {Aoki}}, \bibinfo {author} {\bibfnamefont
			{D.}~\bibnamefont {Braithwaite}}, \bibinfo {author} {\bibfnamefont
			{J.}~\bibnamefont {Flouquet}},\ and\ \bibinfo {author} {\bibfnamefont
			{J.~P.}\ \bibnamefont {Brison}},\ }\bibfield  {title} {\bibinfo {title}
		{{Field-Induced Tuning of the Pairing State in a Superconductor}},\ }\href
	{https://doi.org/10.1103/PhysRevX.13.011022} {\bibfield  {journal} {\bibinfo
			{journal} {Phys. Rev. X}\ }\textbf {\bibinfo {volume} {13}},\ \bibinfo
		{pages} {011022} (\bibinfo {year} {2023})}\BibitemShut {NoStop}%
	\bibitem [{\citenamefont {Wu}\ \emph {et~al.}(2017)\citenamefont {Wu},
		\citenamefont {Bastien}, \citenamefont {Taupin}, \citenamefont {Paulsen},
		\citenamefont {Howald}, \citenamefont {Aoki},\ and\ \citenamefont
		{Brison}}]{Wu2017}%
	\BibitemOpen
	\bibfield  {author} {\bibinfo {author} {\bibfnamefont {B.}~\bibnamefont
			{Wu}}, \bibinfo {author} {\bibfnamefont {G.}~\bibnamefont {Bastien}},
		\bibinfo {author} {\bibfnamefont {M.}~\bibnamefont {Taupin}}, \bibinfo
		{author} {\bibfnamefont {C.}~\bibnamefont {Paulsen}}, \bibinfo {author}
		{\bibfnamefont {L.}~\bibnamefont {Howald}}, \bibinfo {author} {\bibfnamefont
			{D.}~\bibnamefont {Aoki}},\ and\ \bibinfo {author} {\bibfnamefont {J.-P.}\
			\bibnamefont {Brison}},\ }\bibfield  {title} {\bibinfo {title} {{Pairing
				mechanism in the ferromagnetic superconductor UCoGe}},\ }\href
	{https://doi.org/10.1038/ncomms14480} {\bibfield  {journal} {\bibinfo
			{journal} {Nat. Commun.}\ }\textbf {\bibinfo {volume} {8}},\ \bibinfo {pages}
		{14480} (\bibinfo {year} {2017})}\BibitemShut {NoStop}%
\end{thebibliography}

%

\newpage

\onecolumngrid

\section{Supplemental material to the manuscript \\ "Connecting High-Field and High-Pressure Superconductivity in UTe$_2$"}

\section{Experimental details}
\subsection{Setups for measuring}
All the samples of UTe$_2$ of this study were synthesized (either in Grenoble or Oarai) using the recently developed Molten Salt Flux (MSF) technique \cite{Sakai2022}\cite{Aoki2024}. 

\subsubsection{Sample A - LNCMI Grenoble}
Measurements on sample A have been performed at the high magnetic field laboratory in Grenoble. We used a $^3$He cryostat to measure in temperature range from  $\approx800$~mK and up to $\approx3.5$~K. The high pressure cell has been fixed on a silver cold finger attached to the $^3$He stage. The maximal available magnetic field was 30~T. The temperature has been measured with a calibrated RuO$_2$ thermometer which had been glued on the pressure cell. As heater we used a commercial 120~$\Omega$ strain gauge which has been also attached to the pressure cell. 

The measurements under pressure have been performed with a piston cylinder CuBe/NiCrAl pressure cell with inner diameter of $5$~mm and outer diameter of $20$~mm. As pressure medium we used Daphne oil 7373. 
To determine the pressure inside the cell, we used a manganin coil, as well as a miniature ac susceptibility setup: a copper coil around a Pb wire. At room temperature, we measured the relative change in resistance of the manganin wire to determine the amount of force to apply during loading or when changing the pressure in the cell (Ref.~\cite{xiang_characterization_2020}). To know the pressure at low temperature, we measured the superconducting critical temperature of Pb by ac susceptibility in a Quantum Design PPMS, where we carefully corrected for the remnant field in the superconducting magnet. The pressure has been cross-checked at the LNCMI prior the measurements. 

\subsubsection{Sample B - CEA Grenoble}
Sample B has been grown and studied at the CEA Grenoble. We used a $^3$He/$^4$He dilution fridge to measure Sample B in temperature range down to $\approx800$~mK and up to $4$~K. Maximum fields up to $8$~T were achieved using a superconducting magnet.
We used a diamond anvil pressure cell with Ar as pressure medium. The pressure could be changed in-situ at low temperatures using a $^4$Helium bellows system.  Pressure at low temperature has been measured by the frequency shift of the R1 fluorescence line.

\subsubsection{Sample C - IMR Oarai}
Sample C was measured at IMR Oarai. We used a top-loading $^3$He/$^4$He dilution fridge to measure in temperature range down to $\approx100$~mK and up to $4$~K. Measurements were performed with magnetic fields up to $14.7$~T.

To determine the pressure inside the cell at low temperature, we measured the ac susceptibility of Pb, as for Sample A.\\

\subsubsection{Ac calorimetry setups}
We prepared calorimetry setups, meaning that every sample is coupled to a heater and a thermometer. Three types of heater/thermometer were used in total, and were prepared as such:
\begin{itemize}
	\item \textbf{Carbon paste (heater): }a $15$~µm gold wire is bonded to the sample by spot welding. A small drop of resistive carbon paste (of resistance typically in the range of $10$-$100$~$\Omega$) is then deposited on top the wire and the sample. Finally, another gold wire is simply placed upon the drop, and the whole setup is set at 150$^{\circ}$C for an hour for the resin to polymerize.
	\item \textbf{Cernox (thermometer): }the temperature of the sample is measured using a Cernox thermometer (following a method similar to Ref.~\cite{gati_use_2019}). The set sample+heater is simply glued to the thermometer with a drop of DuPont 4929N silver paste.
	\item \textbf{Thermocouple (heater/thermometer): }a Au wire and a AuFe wire are flattened, then bonded together directly on the sample. This can be used both as a heating device, or a thermometer.
\end{itemize}
Details on which setup was used on each sample is provided, as well as a recap on other experimental aspects, are given in Table~\ref{tab:recap_table}.
\subsection{Detailed measurement process for Sample A}
When performing the calorimetry measurements, we excite the sample with a sinusoidal current of frequency $f$ through the heater. We then pass a DC current through the thermometer, measuring both the DC resistance $R_{DC}$ (corresponding to a constant temperature response $T_{DC}$), as well as the double-harmonic AC resistance oscillations $R_{AC}$. The associated temperature oscillations amplitude is then given by:
\[ \left\vert \underline{T_{AC}} \right\vert = R_{AC}\left.\frac{\partial T}{\partial R}\right\vert_{T=T_{DC}},\]
, where the slope of the Cernox characteristic is determined with a $\log R$ - $\log T$ polynomial fit for T-sweeps, and approximated by a 2D $T$-$H$ interpolation model for H sweeps.\\

Since we are able to measure the resistance of the heater, we can reconstruct the temperature dependence of the heating power $Q(T)$. Additionally, a frequency analysis led us to choose a measurement frequency of $f_0=223$~Hz to decouple the sample from the environment, and showed that our circuits induce a constant dephasing term of $\phi = -10^{\circ}$, in the phase $\theta= \text{Arg}(\underline{T_{AC}})$. Finally, we retrieve the specific heat using both the amplitude and the phase of the thermometer double-harmonic response:
\[ C(T) = -Q(T) \frac{\text{sin}(\theta + \phi)}{4\pi f_0\left\vert \underline{T_{AC}}\right\vert}\]

\begin{table*}[ht!]
	\centering
	\begin{tabular}{|c|c||c||c|}
		\cline{2-4}
		\multicolumn{1}{c|}{} & \textbf{Sample A} & \textbf{Sample B} & \textbf{Sample C}\\ \cline{2-4}\cline{1-4}
		\textbf{Measured in} & LNCMI & CEA Grenoble & Ōarai\\\hline
		\textbf{Synthesis method} & MSF & MSF & MSF\\\hline
		\textbf{Zero-field} $\boldsymbol{T_c}$ & $\approx1.95$~K & $\approx1.8$~K & $\approx2.1$~K\\\hline\hline
		\textbf{Maximum field} & 30~T & 8~T & 14.7~T\\\hline
		\textbf{Cryogenic setup} & \textsuperscript{3}He cryostat & \textsuperscript{3}He/\textsuperscript{4}He dilution fridge & \textsuperscript{3}He/\textsuperscript{4}He dilution fridge \\\hline\hline
		\textbf{Pressure cell} & piston-cylinder & diamond anvil cell & piston-cylinder\\\hline
		\textbf{Heater} & carbon paste & carbon paste & Au/AuFe thermocouple\\\hline
		\textbf{Thermometer} & Cernox & Au/AuFe thermocouple & Au/AuFe thermocouple\\\hline
		\textbf{Pressure medium} & DAPHNE oil 7373 & Argon & DAPHNE oil 7373\\\hline
	\end{tabular}
	\caption{Summary of the three experiments carried out in this work}
	\label{tab:recap_table}
\end{table*}
\newpage
\section{Specific heat measurements in LNCMI Grenoble}
\begin{figure*}[h!]
	\centering
	\includegraphics[width=1\linewidth]{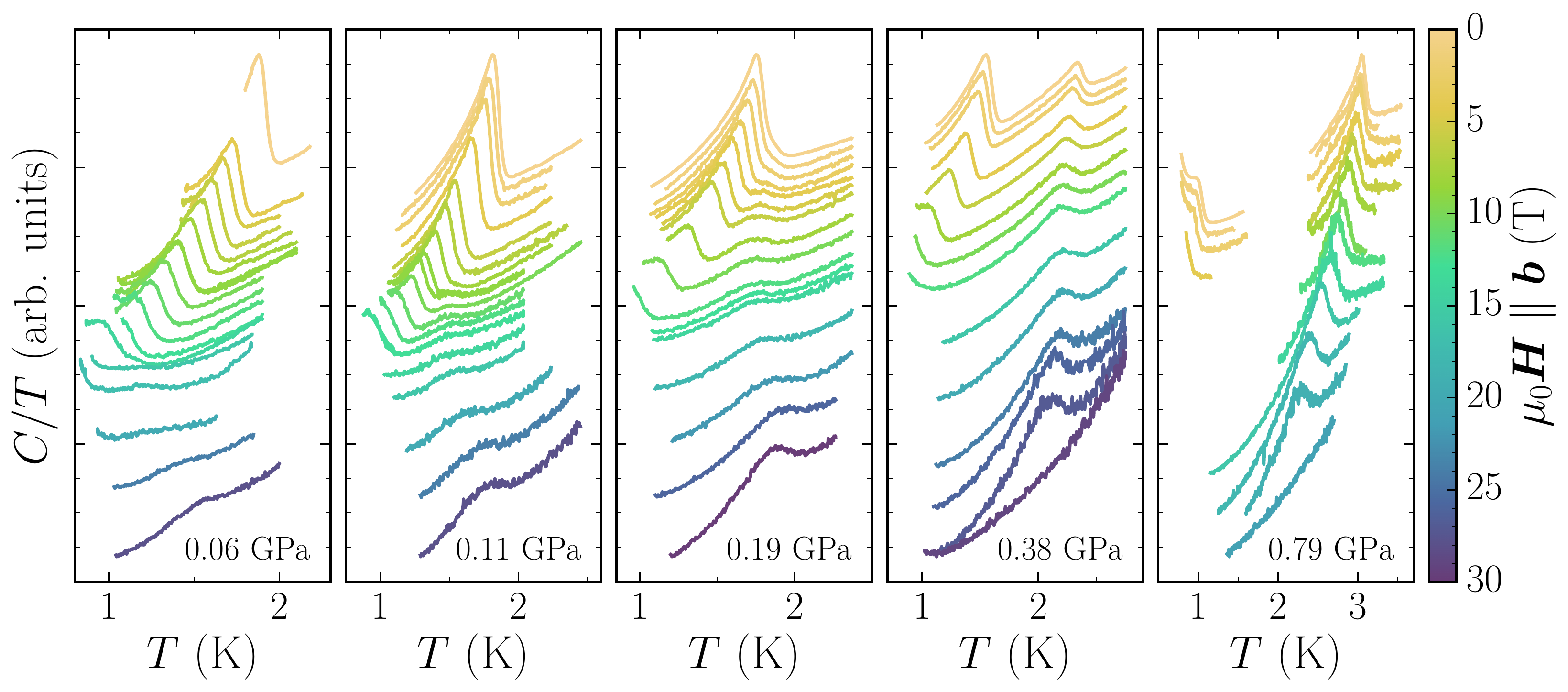}
	\caption{C/T curves at all five pressures measured for Sample A. Each colored curve represent one value for the magnetic field (see colormap on the right). All curves have been vertically shifted for clarity.}
	\label{fig:LNCMI}
\end{figure*}
\section{Specific heat measurements in CEA Grenoble}
\begin{figure*}[h!]
	\centering
	\includegraphics[width=0.465\linewidth]{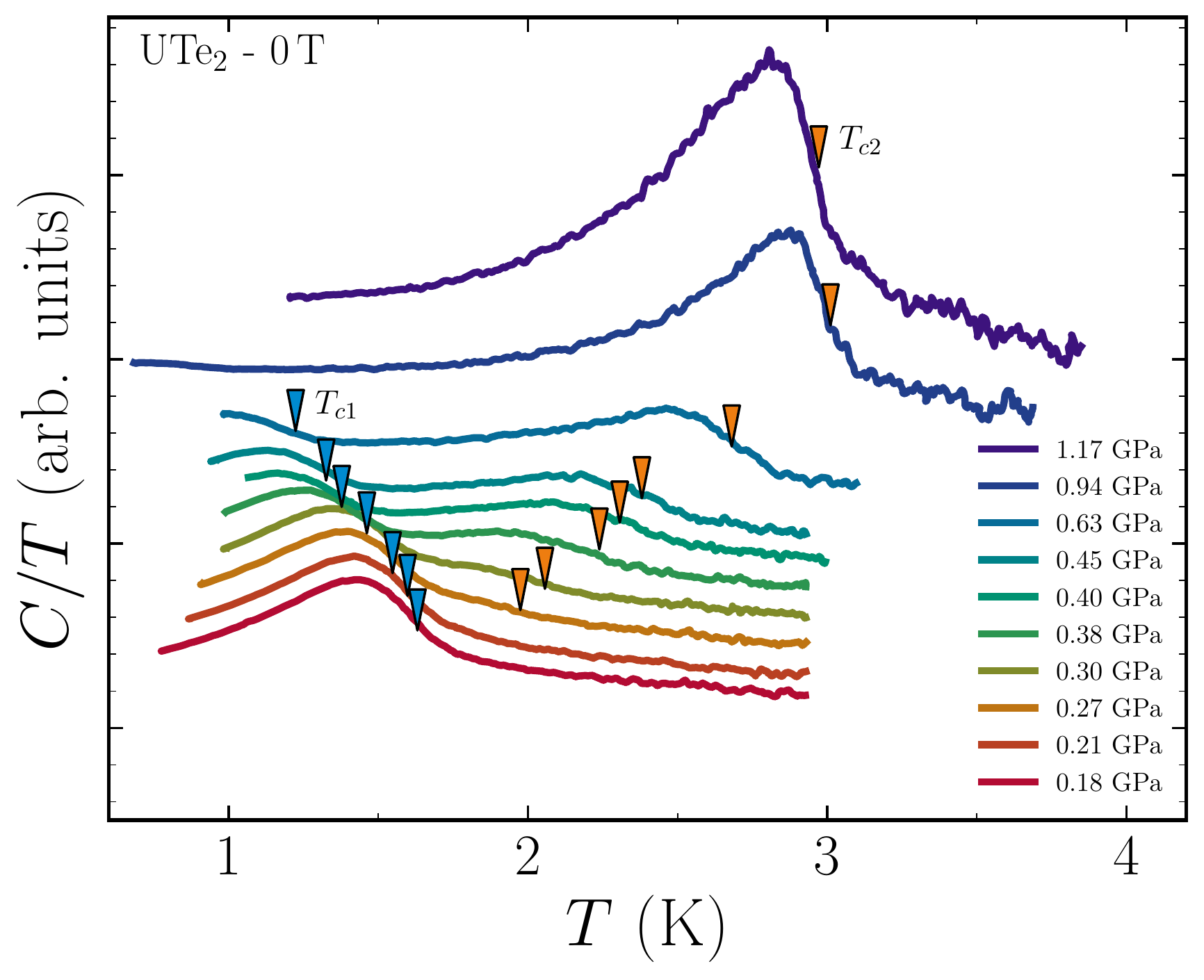}
	\includegraphics[width=0.485\linewidth]{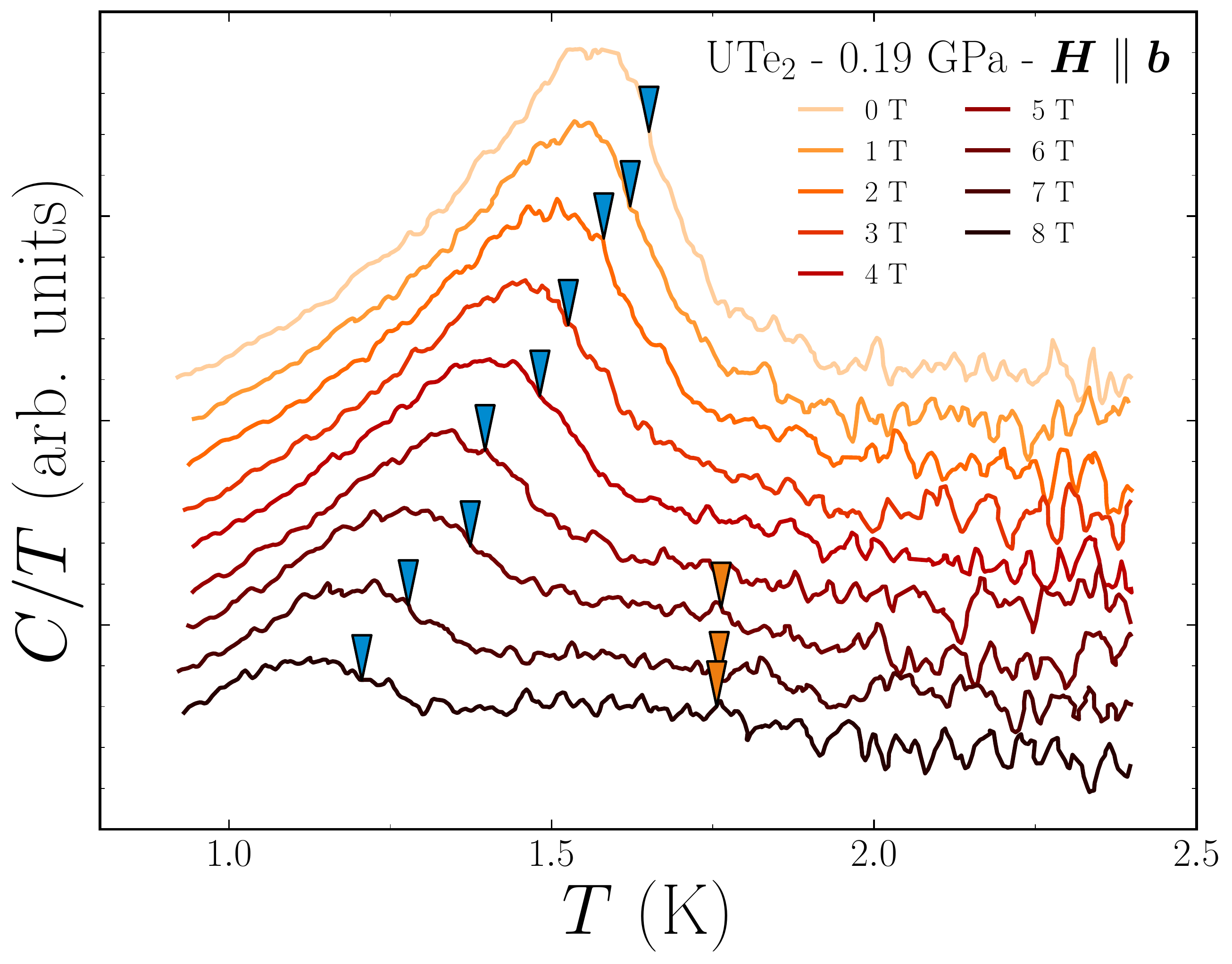}
	\caption{Left: Specific heat measurements for Sample B, at zero field for different pressures. The blue (orange) arrows show the SC1 (SC2) superconducting transition. Right: At a given pressure of $0.19$~GPa, the field-dependence of the specific heat curves is shown. The blue and orange dots show the two superconducting transitions.}
	\label{fig:DAC}
\end{figure*}
\newpage
\section{Specific heat measurements in Ōarai}
\begin{figure*}[ht!]
	\centering
	\includegraphics[width=\textwidth]{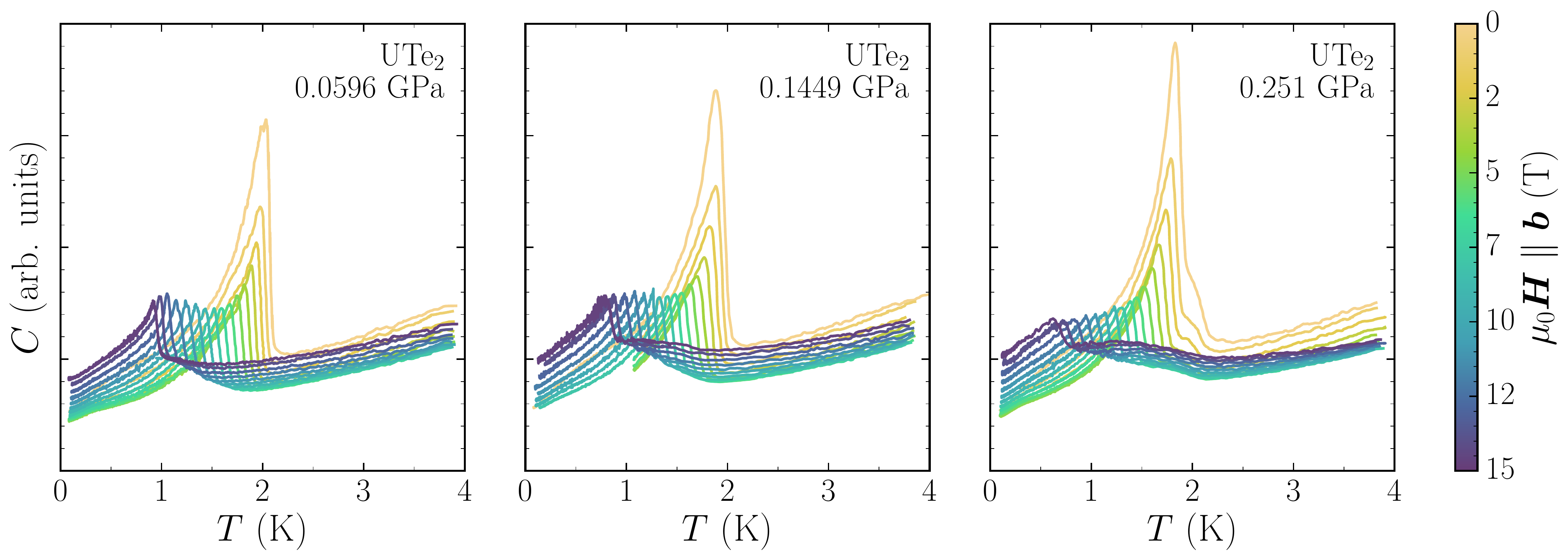}
	\caption{C(T) curves obtained for Sample C at three different pressures. Each colored curve represent one value for the magnetic field (see colormap on the right).}
	\label{fig:dai}
\end{figure*}
\begin{figure*}[ht!]
	\centering
	\includegraphics[width=\textwidth]{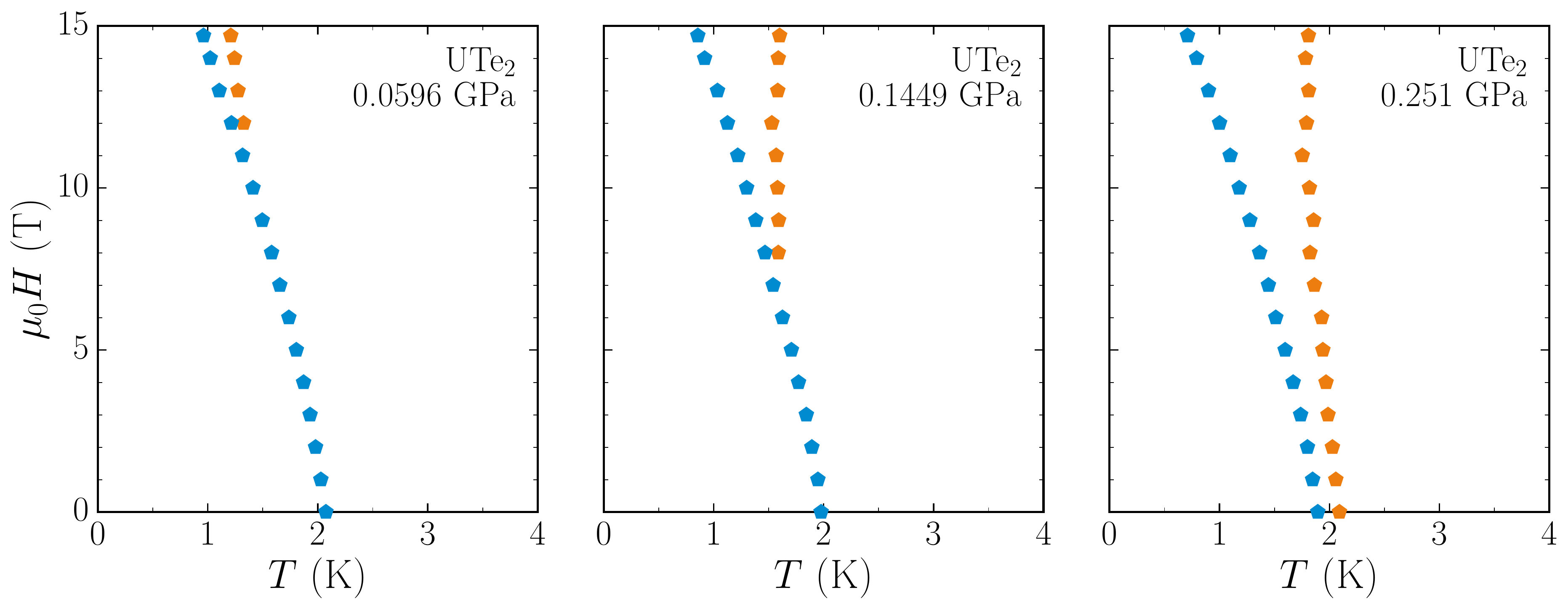}
	\caption{$T$-$H$ phase diagrams of Sample C at three different pressures (see Fig.~\ref{fig:dai} above for specific heat data), for fields up to $14.7$~T. Blue (orange) pentagons represent SC1 (SC2) superconducting transition. Each $T_c$ is placed by eye using the corresponding specific heat curve.}
	\label{fig:dai_ht}
\end{figure*}
\section{Fitting the superconducting jumps}
To consistently apply the same method to place points on our phase diagrams, as well as retrieving additional information such as the transition width, we used a quantitative fitting method adapted from Ref.\cite{Rosuel2023}. To the end of extracting the superconducting critical temperature from specific heat data, we start from a very simple model of an ideal, infinitely sharp transition for a given $T_c$:
\[ C/T^{id}= \underbrace{{\cal{H}}(T_c-T)}_{\text{Heaviside step}}\overbrace{\left(\Delta  C/T+\alpha(T-T_c)\right)}^{\text{SC state}} + \overbrace{\gamma + \text{background term}}^{\text{normal state}} \]
where $T_c$ is the critical temperature, $\Delta  C/T$ is the specific heat jump at the transition, $\alpha$ is the slope (assumed constant) below the transition, and $\gamma T$ is the electronic specific heat in the normal state. Indeed, the drawback of this simple model is that it will only apply to a short temperature range around the transition, one of the reason being that the assumption of constant slope below the transition is almost never true on wider temperature ranges.\\

In the case of ac calorimetry under pressure, we often observe a signal background that does not necessarily reflect a true physically relevant phenomenon. In our case, to facilitate the convergence of the fit, we introduced a linear background term $C_{bg}/T=\eta T$ in the model.\\

Now to account for the broadening of the transition, we suppose that the critical temperature in a region of the sample is a random variable $\boldsymbol{T_c}$ following a Gaussian distribution ${\cal{N}}(T_c^0, \sigma)$, controlled by the following probability density:
\[p(\boldsymbol{T_c} = T_c) = \frac{1}{\sigma\sqrt{2\pi}}\exp\left(-\frac{1}{2}\left(\frac{T_c - T_c^0}{\sigma}\right)^{2}\right)\]

Finally, to reconstruct the broadened specific heat, we take a superposition of each possible outcome of $\boldsymbol{T_c}$, weighted by its corresponding probability density. This is essentially a convolution which is expressed as:
\begin{align*}
	C/T(T) &= C/T^{ideal}\ast p=\int_{-\infty}^{+\infty}C/T^{ideal}(T, T_c)p(T_c)\mathrm{d}T_c\\
	&= \gamma + \eta T + \frac{1}{2}\left(1-\text{erf}\left(\frac{T-T_c^0}{\sigma\sqrt{2}}\right)\right)\left(\Delta C/T + \alpha(T-T_c^0)\right) - \frac{\alpha\sigma}{\sqrt{2\pi}}\exp\left(-\frac{1}{2}\left(\frac{T-T_c^0}{\sigma}\right)^2\right)
\end{align*}
Fitting this expression to our data will yield the critical temperature $T_c = T_c^0$, the transition jump $\Delta C/T$, as well as $\sigma$, which represents the width of the transition (apparent transition width $\Delta T_c \approx 2.5\sigma$).\\

For the case of double transitions, we run this fit separately on each jump, but only when the two transitions have a negligible overlap. When the two critical temperatures are too close to each other, we adapt the model to account for a double transition, simply by considering two independent critical temperatures $\boldsymbol{T_{c1}}$ and $\boldsymbol{T_{c2}}$, each following its own Gaussian distribution $\boldsymbol{T_{ci}}\sim{\cal{N}}(T_{ci}^0, \sigma_i)$. We then simply take an affine combination of the two resulting expressions, with $\xi$ as a mixing parameter, giving for the specific heat: $C/T(T) = C/T^{ideal}\ast (\xi p_1 + (1-\xi)p_2)$

\section{Specific heat anomaly}

\begin{figure}[t]
	\includegraphics[width=0.6\linewidth]{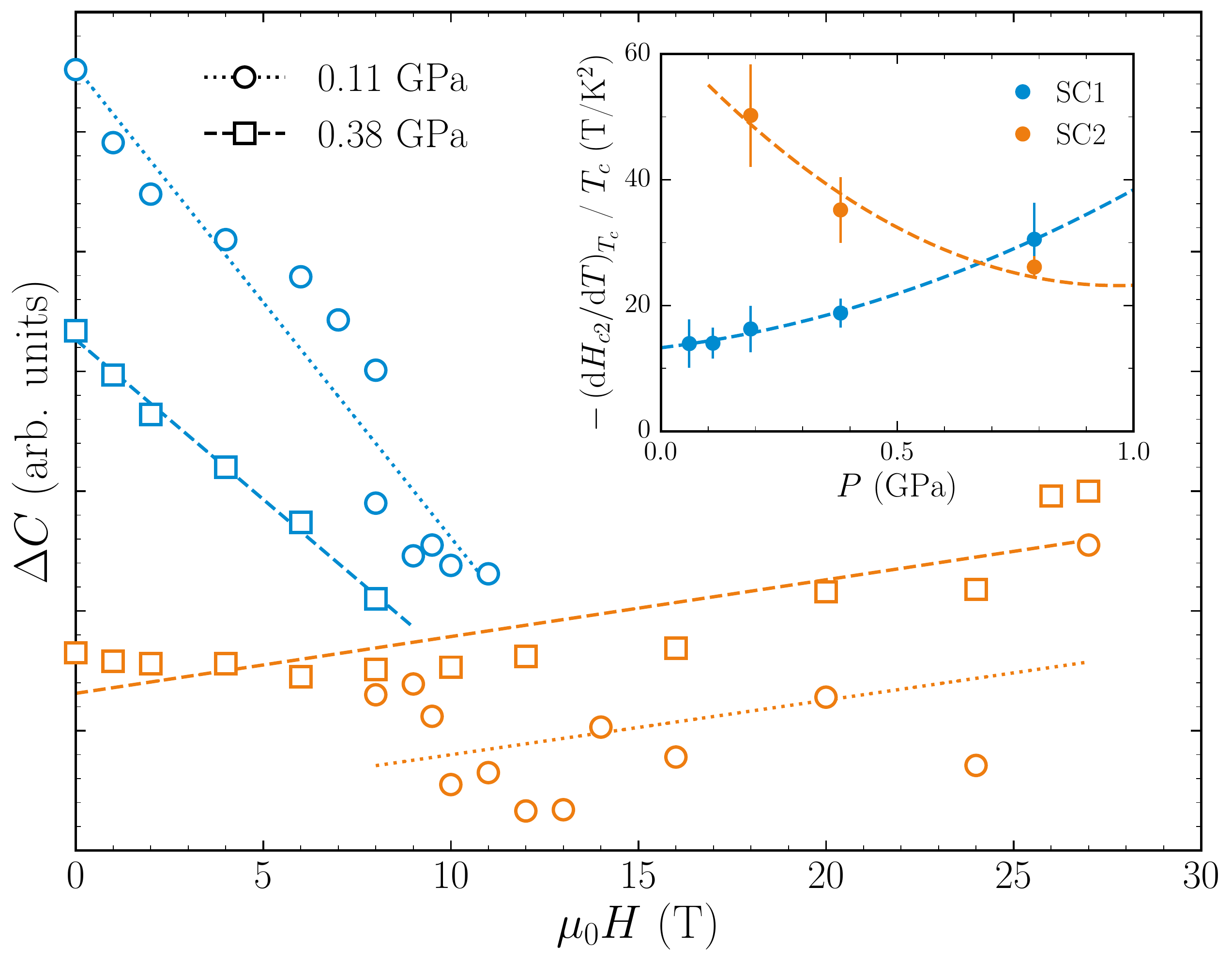}
	\caption{Superconducting specific heat jump $\Delta C/T$ extracted from the Gaussian model, for both transitions at two different pressures. The inset shows the initial slope normalized by $T_c$ for both transitions. (Dashed lines are guides for the eye.)}
	\label{jump}
\end{figure}

In the lower part of Fig.~\ref{jump} we show the height of the jump $\Delta C/T$ for the transitions at $T_{c1}$ and $T_{c2}$ of the measurements on sample A at two pressures (similar to data shown in Fig.~4 of the main article). The height of the jump for SC2 is rather small at low pressures, and shows no pronounced field dependence, similar to the results found in the previous ambient pressure study \cite{Rosuel2023}. The height of this jump increases significantly at higher pressures.
In the inset of Fig.~\ref{jump} we also show the initial slope of the critical field for both superconducting phases, normalized by $T_c$. The initial slope for SC2, at the lowest pressure where it can be determined, is extremely large. 
This is certainly in part due to the effect of reinforcement of superconductivity by the applied field. 
Indeed, at slightly higher fields the slope even becomes positive. 
As pressure increases, the upper critical field 
of SC2 appears to become more normal, and while the initial slope remains extremely large, 
the value normalized to $T_{c2}$ decreases quite significantly.

Indeed, the normalized initial slope of $H_{c2}$ at $T_c$ is controlled both by the pairing strength, leading to an increase until the maximum $T_c$ is reached, and by the field variation of the pairing strength at $T_c$ \cite{Wu2017}, which may decrease on approaching the optimum critical temperature.
This last point is somewhat consistent with the sharpening of the transition at high pressure and low field (main panel of Fig.~4 of the main paper), if the broadening of the transition at SC2 originates from the field dependence of the pairing strength \cite{Rosuel2023}.

We also show the initial slope for SC1. 
The nature of this transition changes with pressure: at low pressure it is a transition between the normal and superconducting states, whereas above $0.2$~GPa it is a transition between two superconducting states. 
Nevertheless, the slope evolves continuously, and when normalized by $T_{c1}$, it actually becomes rather steep.

\section{Transition in the signal phase}
\begin{figure}[t]
	\includegraphics[width=0.54\linewidth]{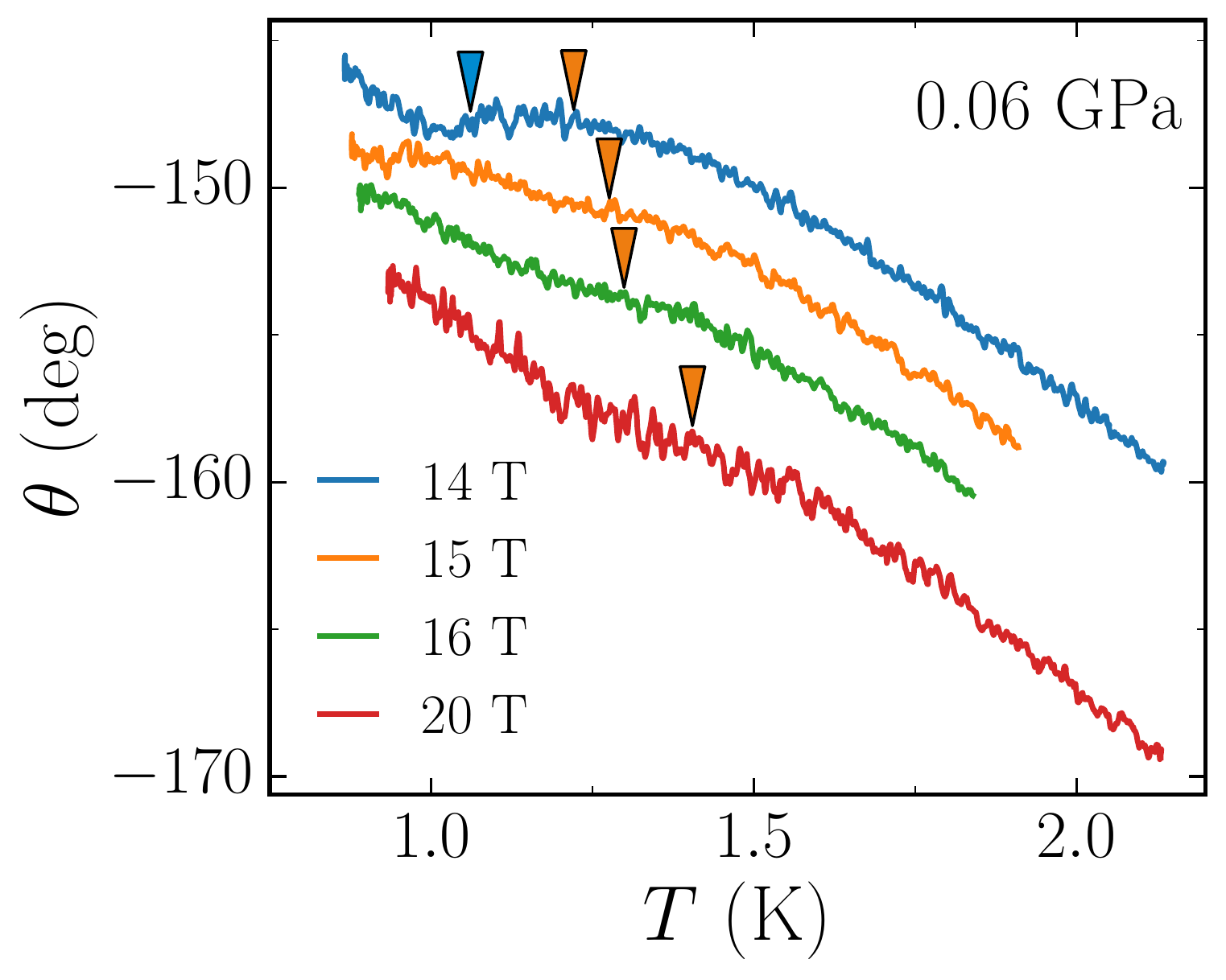}
	\caption{Phase $\theta= \text{Arg}(\underline{T_{AC}})$ of the signal for different fields, at $p=0.06$~GPa. The blue (orange) triangles indicate the superconducting SC1 (SC2) transition, as the midpoint of the transition in the phase. These points are shown in in grey in the main paper.}
	\label{phase}
\end{figure}

Whenever the overlap between the two superconducting phase transitions was too important, and when the transition jump at SC2 was too small, the Gaussian fitting technique could not properly resolve the shape of the transition. In that case, we placed the transition temperatures by eye, at the middle of the transition in the phase (see Fig.~\ref{phase})

apsrev4-2.bst 2019-01-14 (MD) hand-edited version of apsrev4-1.bst

\end{document}